\documentclass[twocolumn]{aastex631}

\newcommand{\orcid}[1]{\href{https://orcid.org/#1}
{\includegraphics[width=8pt]{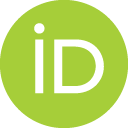}}}

\usepackage{verbatim}

\usepackage{amsmath}

\submitjournal{ApJL}

\begin{document}

\title{JWST unveils a high mean molecular weight atmosphere for mini-Neptune TOI-1130b: Evidence for formation beyond the water ice line}


\footnotetext{This study uses CHEOPS data observed as part of the Guaranteed Time Observation (GTO) programme CH\_PR149003.}

\correspondingauthor{Saugata Barat}
\email{saugatabarat500@gmail.com}

\author[0009-0000-6113-0157]{Saugata Barat}

\affiliation{Kavli Institute for Astrophysics and Space Research, Massachusetts Institute of Technology, Cambridge, MA 02139, USA}

\affiliation{University of Southern Queensland, West St, Darling Heights, Toowoomba,
Queensland, 4350, Australia}

\author[0000-0002-0692-7822]{Tyler Fairnington}
\affiliation{University of Southern Queensland, West St, Darling Heights, Toowoomba,
Queensland, 4350, Australia}
\affiliation{Department of Astronomy \& Astrophysics, University of Chicago, Chicago, IL 60637, USA}

\author[0009-0002-9017-3679]{Shelby Courreges}
\affiliation{Department of Astronomy, University of Texas, Austin, TX 78712, USA}

\author[0000-0003-0918-7484]{Chelsea Huang}
\affiliation{University of Southern Queensland, West St, Darling Heights, Toowoomba,
Queensland, 4350, Australia}

\author[0000-0001-7246-5438]{Andrew Vanderburg}
\affiliation{Kavli Institute for Astrophysics and Space Research, Massachusetts Institute of Technology, Cambridge, MA 02139, USA}

\affiliation{Center for Astrophysics — Harvard and Smithsonian, 60 Garden Street, Cambridge, MA 02138, USA}

\affiliation{Sloan Research Fellow}

\author[0000-0002-4404-0456]{Caroline V. Morley}
\affiliation{Department of Astronomy, University of Texas, Austin, TX 78712, USA}

\author[0000-0002-0076-6239]{Judith Korth}
\affiliation{Observatoire astronomique de l’Université de Genève, Chemin Pegasi 51, 1290 Versoix, Switzerland}
\affiliation{Lund Observatory, Division of Astrophysics, Department of Physics, Lund University, Box 118, 22100 Lund, Sweden}

\author[0000-0001-5519-1391]{Hannu Parviainen}
\affiliation{Departamento de Astrof\'isica, Universidad de La Laguna (ULL), E-38206 La Laguna, Tenerife, Spain}
\affiliation{Instituto de Astrof´ısica de Canarias (IAC), E-38200 La Laguna, Tenerife, Spain}

\author[0000-0002-7201-7536]{Alexis Brandeker}
\affiliation{Department of Astronomy, Stockholm University, AlbaNova University Center, 10691 Stockholm, Sweden}

\author[0000-0002-4891-3517]{George Zhou}
\affiliation{University of Southern Queensland, West St, Darling Heights, Toowoomba,
Queensland, 4350, Australia}

\author[0000-0001-5442-1300]{Thomas M. Evans-Soma}
\affiliation{School of Science, University of Newcastle, Callaghan, NSW, Australia}

\author[0000-0001-5401-8079]{Lizhou Sha}
\affiliation{Department of Astrophysical Sciences, Princeton University, 4 Ivy Ln, Princeton, NJ 08540 USA}

\author[0000-0001-5466-4628]{Douglas N. C. Lin }
\affiliation{Department of Astronomy and Astrophysics, University of California, Santa Cruz, CA 95064, USA}
\affiliation{Institute for Advanced Studies, Tsinghua University, Beijing, 100086, China}

\author[0000-0001-7294-5386]{Duncan Wright}
\affiliation{University of Southern Queensland, West St, Darling Heights, Toowoomba,
Queensland, 4350, Australia}

\author[0009-0000-3527-8860]{Ava Morrissey}
\affiliation{Carnegie Science Observatories, 813 Santa Barbara Street,
Pasadena, CA91101, USA}

\author[0000-0003-0571-2245]{Emma Nabbie}
\affiliation{University of Southern Queensland, West St, Darling Heights, Toowoomba,
Queensland, 4350, Australia}

\author[0000-0001-6588-9574]{Karen A.\ Collins}
\affiliation{Center for Astrophysics \textbar \ Harvard \& Smithsonian, 60 Garden Street, Cambridge, MA 02138, USA}

\author[0000-0002-5674-2404]{Phil Evans}
\affiliation{Phil Evans, El Sauce Observatory, Coquimbo Province, Chile}

\author[0000-0002-7188-8428]{Tristan Guillot}
\affiliation{Universit\'e C\^ote d'Azur, Observatoire de la C\^ote d'Azur, CNRS, Laboratoire Lagrange, Bd de l'Observatoire, CS 34229, 06304 Nice cedex 4, France}

\author[0000-0003-1728-0304]{Keith Horne}
\affiliation{SUPA School of Physics and Astronomy, University of St\,Andrews, North Haugh, St\,Andrews KY16\,9SS Scotland, UK}

\author{Don J. Radford}
\affiliation{Brierfield Observatory, Bowral, NSW, Australia}

\author[0000-0001-8227-1020]{Richard P. Schwarz}
\affiliation{Center for Astrophysics \textbar \ Harvard \& Smithsonian, 60 Garden Street, Cambridge, MA 02138, USA}

\author[0000-0002-1836-3120]{Avi Shporer}
\affiliation{Department of Physics and Kavli Institute for Astrophysics and Space Research, Massachusetts Institute of Technology, Cambridge, MA 02139, USA}

\author{Gregorg Srdoc}
\affil{Kotizarovci Observatory, Sarsoni 90, 51216 Viskovo, Croatia}

\author{Olga Suarez}
\affiliation{Universit\'e C\^ote d'Azur, Observatoire de la C\^ote d'Azur, CNRS, Laboratoire Lagrange, Bd de l'Observatoire, CS 34229, 06304 Nice cedex 4, France}
abstract in the ``abstract'' environment. 
\begin{abstract}

We present the combined JWST/NIRSpec G395H and NIRISS SOSS transmission spectrum of a warm mini-Neptune, TOI-1130b (3.66~R$_{\oplus}$, 19.8~M$_{\oplus}$, $T_\mathrm{eq}\sim825$~K). It is part of a rare and unique multi-planet system TOI-1130, which hosts an inner mini-Neptune and an outer hot Jupiter locked in a 2:1 mean motion resonance. From the transmission spectrum of TOI-1130b we detect multiple molecules --- H$_2$O (7.5$\sigma$), CO$_2$ (3.3$\sigma$), and SO$_2$ (3.6$\sigma$), as well as a tentative detection of CH$_4$ ($\sim$2$\sigma$). We find a strong optical slope in the NIRISS/SOSS spectrum, which is consistent with TESS and CHEOPS transit depth measurements. From equilibrium chemistry retrievals we measure the atmospheric metallicity ($\log{Z/Z_{\odot}}=1.8^{+0.4}_{-0.3}$) and C/O ratio ($<$0.75 at 3$\sigma$ level confidence) and constrain the atmospheric mean molecular weight, $\mu$ = 5.5$^{+1.3}_{-0.8}$ amu. These constraints are consistent with self-consistent forward model grids. We detect no significant He\,I\,1.083\,$\mu$m absorption signal and put a mass-loss rate upper limit of $10^{11}$g\,s$^{-1}$.
The volatile-rich high mean molecular weight atmosphere of TOI-1130b along with the `pebble-filtering' effect of the outer hot Jupiter supports the ex-situ formation scenario beyond the water ice line and subsequent migration, coherent with its present orbital architecture. A volatile-rich formation scenario could also potentially explain the location of TOI-1130b at the edge of the `radius cliff'. This result hints that the mini-Neptune population may not a homogeneous formation history; rather, volatile-rich ex-situ formation also contributes to its population.

\end{abstract}


\section{Introduction}

The first detection of a hot Jupiter, 51 Pegasi b \citep{mayor1995} occurred more than three decades ago, yet we do not understand their formation history \citep{dawson2018,fortney2021}. Several theories have been suggested: formation beyond the ice line and inward migration due to high eccentricity migration \citep{wu2003,chatterjee2008,naoz2016} or disk-driven migration \citep{lin1996,ida_lin2008} as well as in-situ formation \citep{batygin2016}. However, assembling a large core to facilitate core accretion \citep{pollack1996} close to the host star is challenging due to lack of available solids \citep{sclichting2014,lee_chiang2016,johansen2017}.

It has been proposed that atmospheric chemical composition can be used to infer formation location within the protoplanetary disk \citep{oberg_2011,khorshid2022,penzlin2024}, and thus test planet formation theories. However, several degeneracies exist, such as uncertainty about disk chemical composition and planet formation processes (\citealt{molliere2022,feinstein2025} and references therein). Therefore, it is necessary to calibrate the correlation between planet formation location and atmospheric chemical composition.

We know of certain systems for which formation history is well constrained. For example, a handful of hot Jupiters ($<$10 day period) have been discovered which have inner companions \citep{becker2015,huang2020,sha2023}. Typically, hot Jupiters are not found with inner companions \citep{huang2016}, and this has been interpreted as evidence for high eccentricity migration where inner planets are scattered out of the system \citep{mustill2015,dawson2018}. Therefore, multiplanet systems with an inner planet and an outer hot Jupiter are rare: only $\sim$ 7\% of systems with hot Jupiters have inner companions \citep{sha2026}, highlighting their rarity. These systems are most likely to be a product of formation through disk-driven migration. These systems with their well constrained formation history, are ideally suited to test the correlation between planet formation location and atmospheric composition.

In this paper we present the first results from the JWST program GO3385 which aims to address the questions outlined in the last paragraph by characterizing and comparing the atmospheres of two planets in the TOI-1130 system \citep{huang2020}. TOI-1130 is a multi-transiting system with an inner mini-Neptune sized planet, TOI-1130b (3.66$\pm$0.04~R$_{\oplus}$, 19.8$\pm$0.03~M$_{\oplus}$, 4.07 day period), and an outer hot Jupiter TOI-1130c (13$\pm$0.4~R$_{\oplus}$, 336$\pm$5~M$_{\oplus}$, 8.35 day period). The parameters of the system have been determined using photodynamical models \citep{borsato2024,korth2023}. The TOI-1130 system is in a 2:1 mean motion resonance and exhibits transit timing variations (TTVs) of the order of 2 hours \citep{korth2023}. This indicates that the system likely formed through migration from beyond the ice line \citep{beague2006} which results in the capture of the two planets in the 2:1 resonance configuration. Therefore, the orbital configuration and resonance strongly favors formation beyond the ice line and inward migration driven by the disk for both planets.

This paper particularly focuses on the inner mini-Neptune, TOI-1130b. It is a warm mini-Neptune (825~K equilibrium temperature assuming albedo of 0.1) lying at the edge of the radius cliff. The radius cliff is a feature in the radius-period diagram where the occurrence rate drops off sharply for planets larger than 3~R$_{\oplus}$ \citep{fulton2017,Hsu2019,Dattilo2023} with orbital period smaller than 100 days. The existence of the radius cliff is predicted by several early evolution models including photoevaporation \citep{owen_jackson2012,owen2013,owen2017},  core-powered mass loss \citep{ginzburg2018,gupta2019,owen_schlichting2024} and/or boil-off \citep{owen2016}. However, recent studies have shown that atmospheric escape processes alone cannot adequately explain the observed shape of the cliff  \citep{dattilo2024} and additional factors such as different formation pathways \citep{burn2024,chakraborty2026} or core-mass distributions \citep{lee2025} may play a role. TOI-1130b lies at a relatively rare part of the period-radius parameter space, which makes the TOI-1130 with its already unique architecture, a one-of-a-kind system. The presence of the outer hot Jupiter beyond the orbit of TOI-1130b ensures that inflowing volatile-rich material from beyond its orbit is blocked due to `pebble-filtering' by the hot Jupiter \citep{bitsch2021,schneider2021}. Although the efficiency of pebble filtering depends on the mass of the giant planet and diffusion strength \citep{zhu2012,bitsch2018}, for solar system models with proto-Jupiter it has been shown to be efficient \citep{haugbolle2019} and observations of transitional disks have shown that grain growth in the inner disk becomes inefficient once pebbles have been blocked \citep{norfolk2021}. Therefore, we can assume that the chemical composition of the gas/solids which TOI-1130b would accrete is only dependent on the location of accretion within the disk.

The paper is organized as follows: Section\,\ref{observations} presents the observations, Section \ref{section:data analysis} the data analysis,  Section\,\ref{results} the key results and Section\,\ref{discussion} the discussion. We summarize and conclude our interpretations in Section\,\ref{summary}.

\section{Observations} \label{observations}

 \subsection{TESS and CHEOPS}

We used 12 CHEOPS transit observations of TOI-1130b in total. These transits are from the Guaranteed Time Observation of the CHaracterising ExOPplanet Satellite \citep[CHEOPS;][]{2021ExA....51..109B} light curves published by \cite{borsato2024} \footnote{ One of the additional CHEOPS transit from \citet{borsato2024} overlaps with TOI-1130 c, which is excluded from our analysis.}. In addition, CHEOPS observed one transit of TOI-1130~b simultaneously with JWST NIRISS/SOSS transit as part of the Simultaneous exoplanet observations with JWST program with Programme ID: CH\_PR149003 led by Alexis Brandeker. Because TOI-1130 is relatively faint for CHEOPS, we processed the data using the PSF Imagette Photometric Extraction (PIPE) pipeline developed for CHEOPS \citep{2024ascl.soft04002B} as used in \cite{2022A&A...659L...4B} instead of the Data Reduction Pipeline \citep[DRP;][]{2020A&A...635A..24H}. PIPE employs a principal component analysis (PCA) approach to construct a PSF template library from the image series. The individual PSFs in each frame are then fitted with the derived principal components and a constant background term to measure the target flux.

We made use of 20 TESS transits from sectors 13, 27, 67 and 94. We use the highest cadence TESS Presearch Data Conditioning (PDC) light curves \citep{2012PASP..124.1000S,2012PASP..124..985S,2014PASP..126..100S} when available, otherwise we use the QLP FFI light curves \citep{QLP1, QLP2}. Two of the additional TESS transits overlaps with TOI-1130 c. They are also excluded from our analysis. 

\subsection{Refining the transit timing solution}

TOI-1130~b is the only planet observed by JWST with a transit-timing variation amplitude greater than two hours, reaching about five hours from peak to peak. 
This required careful planning of the observations. We revised the photodynamical model published by \citep{korth2023} by adding TESS Sector 67 data and 
new ground-based transit observations from LCO and ASTEP \citep{Guillot+2015}. The updated model was then used to identify JWST observable transit windows while avoiding overlapping events,
and the model's accuracy was validated using ground-based observations obtained before the JWST visit. We also compared our transit predictions with those from \citet{borsato2024}, which include CHEOPS, LCO, ASTEP, and TESS data up to Sector 67, and found the two predictions to agree.

\subsection{JWST Observations}

Two consecutive transits of TOI-1130b were obtained using JWST NIRSpec/G395H (20th August 2024) and JWST/NIRISS SOSS (24th August 2024). The NIRSpec observations were obtained using the default S1600A1 slit, G395H grating/F290LP filter and the SUB2048 subarray. The NRSRAPID readout mode was used and the observations were taken with 8 groups per integrations and a total of 2673 integrations obtained through the 6 hour long visit. For NIRISS SOSS, we used SUBSTRIP256 which covers all spectroscopic orders and we choose the NISRAPID readout mode. We used 2 groups per integration and obtained 1320 integrations. The NIRISS visit was also for 6 hours. The number of groups per integration were chosen to ensure that the detector operated in the linear regime. All integrations recorded a flux lower than 60\% of the saturation limit of the detector. The JWST data presented in this article were obtained from the Mikulski Archive for Space Telescopes (MAST) at the Space Telescope Science Institute. The specific observations analyzed can be accessed via doi: 10.17909/nzdt-ce06.

\section{Data analysis} \label{section:data analysis}

\subsection{JWST data reduction} \label{data reduction}

We begin our analysis from the uncalibrated files (``.uncal") for the NIRSpec and NIRISS/SOSS observations. The NIRSpec/G395H images are analyzed using the \texttt{Eureka!}\ data reduction pipeline \citep{eureka}. \texttt{Eureka!}\ is a widely used and calibrated pipeline which has been applied to multiple NIRSpec/G395H analysis and benchmarked with other standard data reduction pipelines \citep[e.g.\ see][]{May2023,alderson2025,barat2025}. We run the first two data reduction stages from \texttt{Eureka!}\ that wraps the standard \texttt{JWST} pipeline functions. We analyze the calibrated Stage 2 outputs from \texttt{Eureka!}\ using \texttt{Eureka!}\ Stage 3 extraction routines. For detailed discussion of the NIRSpec/G395H data reduction see Appendix\,\ref{NIRSpec_reduction}. We also independently reduced NIRSpec/G395H data using the \texttt{exoTEDRF} data reduction pipeline \citep{radica2024}. The two reductions yield consistent results (Figure\,\ref{fig:exoted_eureka}).

We reduce the NIRISS SOSS images using the \texttt{exoTEDRF} data reduction pipeline. \texttt{exoTEDRF} is a standard pipeline that has been used for reduction of multiple NIRISS SOSS time series observations \citep[e.g.\ see][]{radica2025,davenport2025}. We follow the standard data reduction steps that are outlined in Appendix\,\ref{niriss_reduction}. We note that our observations did not include a F277W image, therefore we did not include the order 0 background contamination flagging step. The 1D spectra are extracted using a box extraction.

\subsection{JWST light curve analysis} \label{JWST data light curve analysis}

\begin{figure*}
    \centering
    \includegraphics[width=1\textwidth]{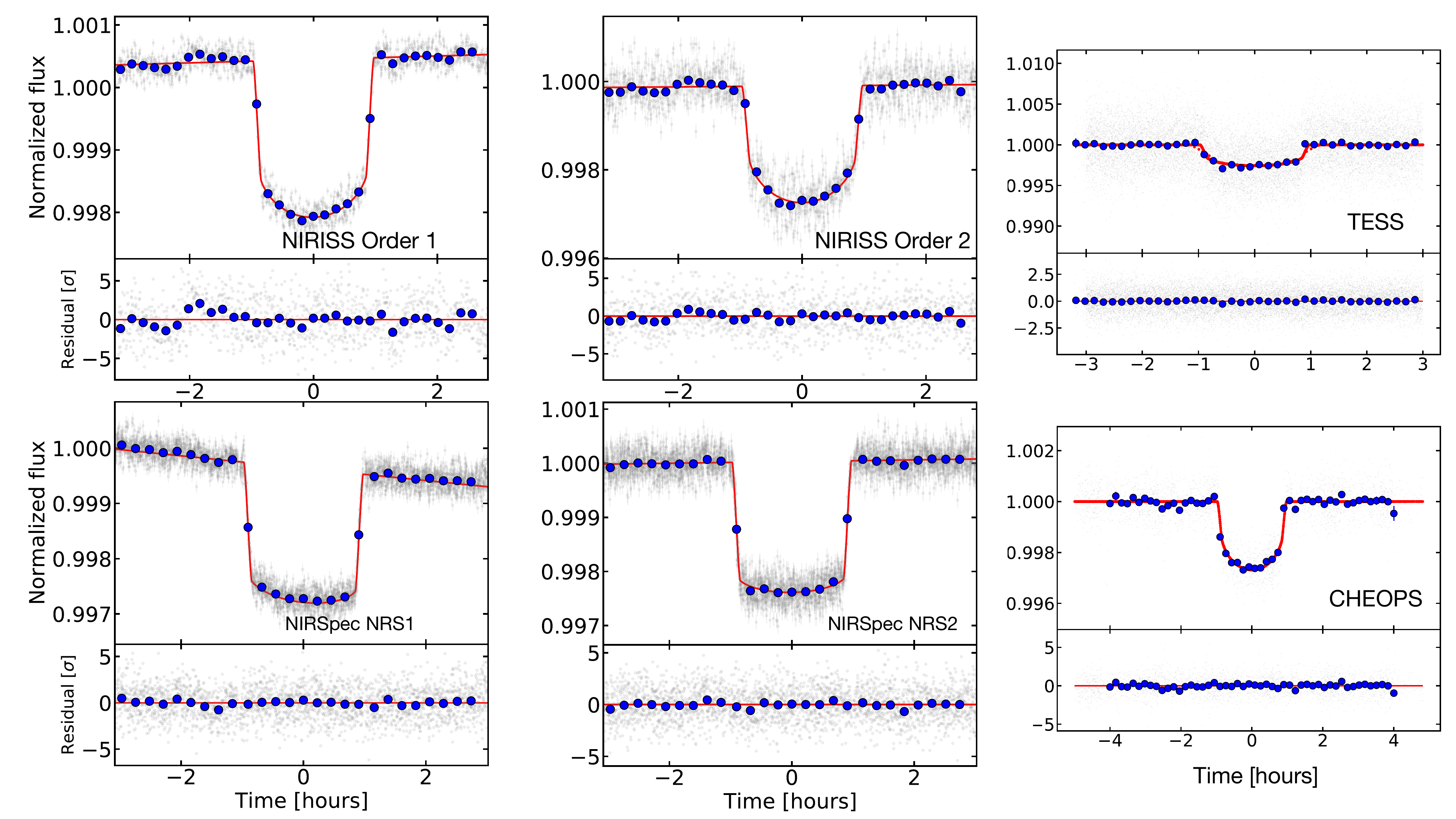}
    \caption{JWST white light curves, phase folded TESS and CHEOPS light curves of TOI-1130b. Upper panel shows NIRISS SOSS (order 1 on the left and order 2 on the right) and lower panel shows NIRSpec G395H light curves (NRS1 on the left and NRS2 on the right). The NIRISS data are reduced using \texttt{exoTEDRF} \citep{radica2025} and the NIRSpec light curves are reduced using \texttt{Eureka!}\ \citep{eureka}. The details of the data reduction are outlined in Section\,\ref{data reduction}. The red solid lines show best-fit models to the white light curves. The white light curve is fit using a \texttt{batman} model with a linear slope and quadratic limb darkening coefficients fixed from \texttt{Exotic-LD} \citep{exotic_ld}. We also show 10 minute-bin light curves (blue circles).  In the NIRISS SOSS light curves we see a `kink' before the transit. In NRS1 we see a visit-long slope that has been seen in other observations \citep{espinoza2023}. NRS2 does not show such a slope. We discuss the light curve fitting methodology in detail in Section\,\ref{JWST data light curve analysis} The right upper panel shows the phase folded TESS and lower right panel shows phase folded CHEOPS light curves. See Section \ref{tess and cheops} for details of analysis.}.
    \label{fig:combined_white_lc}
\end{figure*}

We extract and analyze the transit light curves of TOI-1130b using custom tools. We used the 1D spectra reduced using \texttt{Eureka!}\ and \texttt{exoTEDRF} as discussed in Section\,\ref{data reduction}. First, white light curves are extracted by summing over the full wavelength range of the detectors. The white light curves from NIRSpec/G395H NRS1, NRS2 and NIRISS/SOSS order 1 and order 2 are shown in Figure\,\ref{fig:combined_white_lc}. NRS1 shows a linear slope that has been reported for other similar observations \citep[e.g.][]{espinoza2023,alderson2023}. NRS2 does not show such a slope. The NIRISS/SOSS light curves do not show any apparent slope, but show a `kink' in the pre-transit baseline. Such features have been seen in JWST light curves \citep[e.g.][]{alderson2023} and could be due to mirror tilt/high-gain antenna moves. 

We fit the white transit light curves using a linear baseline and a transit model simulated using \texttt{batman} \citep{batman}. We keep the quadratic limb darkening coefficients, semi-major axis and mid-transit times free for all four white light curves. Other orbital parameters are fixed from \citet{borsato2024}. Further details of the white light curve fitting is described in Appendix\,\ref{subsec:lightcurve_analysis}. 
The residual RMS for the white light curve relative to the expected white noise models from \texttt{Pandexo} \citep{batalha2017} are 2.1, 1.9, 1.1 and 1.04 for NIRISS order 1, order 2, NRS1 and NRS2 respectively. For NIRISS SOSS we have a likely mirror tilt/HGA move which is increasing the RMS. Deviation from expected white noise has been reported for white light curves of multiple other JWST datasets \citep[e.g.][]{wallack2024}, and are consistent with the white noise RMS for the NIRSpec observation.

The spectroscopic light curves are fit at the native resolution of the instruments. We fit the spectroscopic light curves using two approaches: same model as white light curve with no additional red noise model and a common-mode correction approach. The details of these methods are outlined in Appendix\,\ref{subsec:lightcurve_analysis}. In both cases, semi-major axis and mid-transit time are fixed from the respective white light curve fits. We note that the semi-major axis from all the white light curve fits are consistent within 1$\sigma$. The quadratic limb darkening coefficients are fixed from \texttt{Exotic-LD} \citep{exotic_ld}. We note that in the first light curve fitting approach for NIRISS/SOSS light curves, we trim integrations prior to the `kink' in the light curve. From this fit we find that the linear baseline slope is wavelength independent for NRS2, NIRISS order 1 and 2. However, for NRS1 a wavelength dependence is seen that has been reported for previous datasets \citep[e.g.][]{sikora2024}.

In the common-mode approach we use a systematics model derived from the white light curve to detrend the spectroscopic light curves \citep{gibson2013,stevenson2014,kreidberg14}. The systematics model is derived by dividing the observed white light curve with the best-fit transit model. This model assumes that the systematics are wavelength independent and has been applied on many time series observations, especially with HST \citep[e.g.\ see][]{kreidberg14,barat2023}. This approach corrects for the kink in the NIRISS spectroscopic light curves. We include a linear slope as a free parameter in the fitting routine for the detrended spectroscopic light curves. For all the detectors the fitted slope in this approach is consistent with 0, but for NRS1 it shows the expected wavelength dependence. We find that the residuals from the spectroscopic light curve fits (both approaches) bin down similar to expectations from a white noise and the reduced $\chi^{2}$ is consistent with 1 for all the spectroscopic light curve fits. The transmission spectra derived using both these light curve fitting techniques are consistent within 1$\sigma$ for all spectroscopic light curves (Figure\,\ref{fig:normal_common_mode_niriss_spectrum_comparison}). The final transmission spectra are obtained by binning the native resolution transmission spectra. We use 50 pixel bins for NIRISS order 1, NRS1 and NRS2. For NIRISS order 2 we use two spectroscopic channels. The choice of binning is discussed in detail in Appendix\,\ref{subsec:lightcurve_analysis}. The combined JWST transmission spectrum derived from the common-mode correction method for TOI-1130b is shown in Figure\,\ref{fig:full spectrum}. We use this spectrum for the rest of the analysis.

\subsection{TESS and CHEOPS light curve analysis} \label{tess and cheops}

We jointly fit all CHEOPS, TESS data with JWST white light curves from NIRISS/SOSS order 1, NIRSpec NRS1 and NRS2 to provide additional transit depth constraints in the optical. 
We allow different planet radii for different bands, and share the other planet parameters (inclination, semi-major axis of the planet) between the bands. For the CHEOPS and TESS transits, we fix LD values using Exotic-LD similar to the JWST spectra light curve analysis, and fix $t_0$ values for individual transits based on the updated TTV solutions. The JWST white light curves have free LD values, and free $t_0$ values instead. We simultaneously fit the detrending parameters for all data sets. For CHEOPS data, we use detrending sets following the \texttt{pycheops} package \citep{pycheops} with a second order polynomial, for TESS and JWST data, only a linear fit with time is used for detrending. For 30 min and 10 min cadence TESS FFI data, we super sample the transit models before comparing to the observation data.  

We also attempted different combinations of detrending, prior and LD choices for the CHEOPS and TESS data fit. We note the overall TESS transit depth does not change depending on these choices. We find that the CHEOPS transit depths show a maximum variation of $\sim$80\,ppm between visits (typical uncertainties are $\sim$40\,ppm). Therefore, the different CHEOPS visits are consistent within 2$\sigma$ level of significance.

\section{Results} \label{results}

\subsection{Transmission spectrum of TOI-1130b} \label{transmission spectrum}

\begin{figure*}[]
    \centering
    \includegraphics[width=1\textwidth]{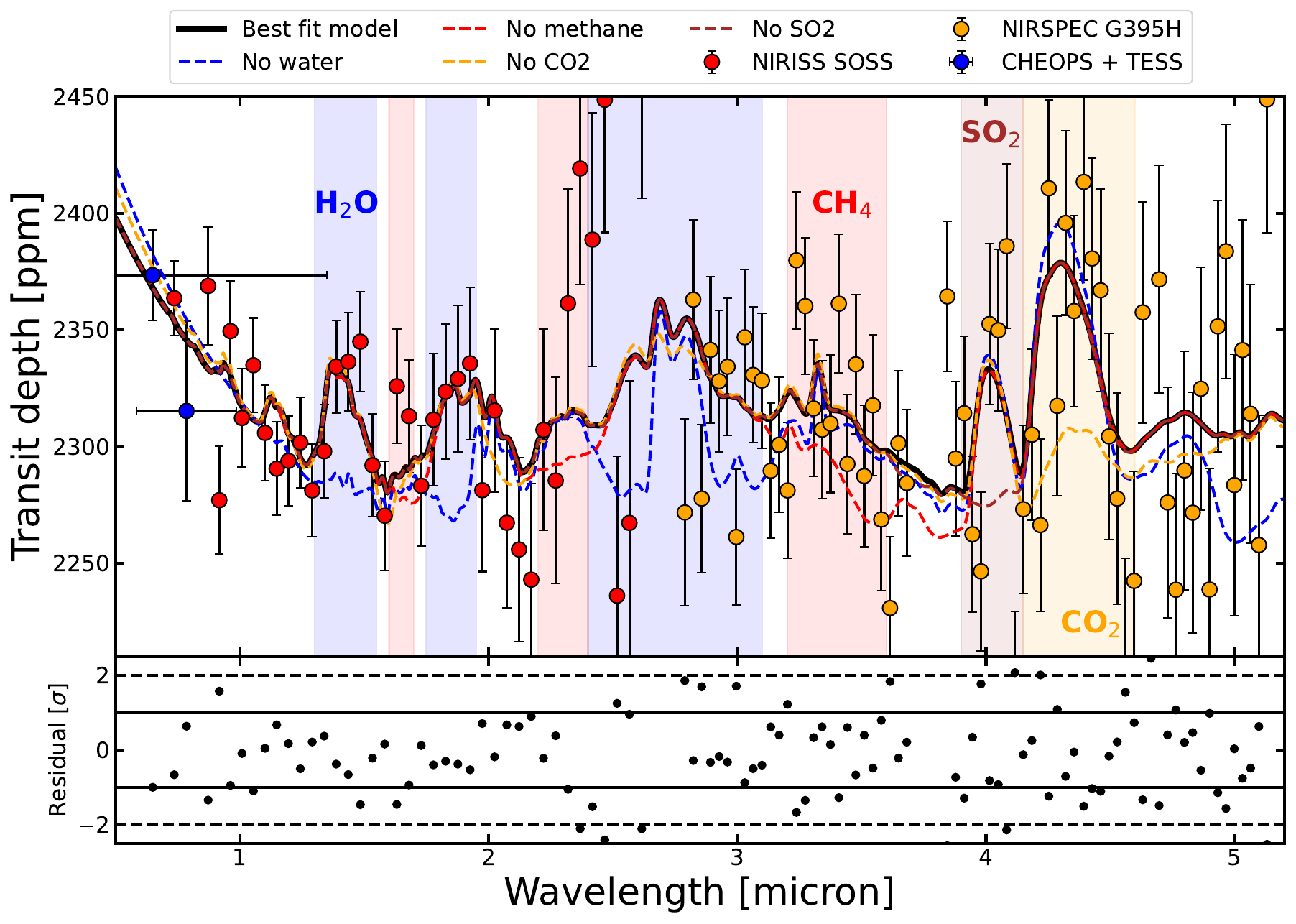}
    \caption{JWST/NIRSpec (orange) and NIRISS SOSS (red) transmission spectrum of TOI-1130b. The blue points show the transit depths derived from TESS and CHEOPS discussed in Section\,\ref{tess and cheops}. The derivation of JWST transmission spectra is discussed in Section\,\ref{JWST data light curve analysis}. The error bars shown have been scaled up by the fitted jitter factor. The black solid model shows the best-fit model from a free chemistry retrieval (including errorbar scaling) implemented using \texttt{PetitRadtrans} \citep{molliere_19}. The best-fit model has a ${\chi}^{2}_{\nu}$=1.01. The details of the forward model are outlined in Section\,\ref{free retrieval}. The different color lines show models where one molecule has been removed from the best-fit model. The vertical dashed regions show the absorption regions corresponding to those molecules. We detect H$_2$O (7.5$\sigma$), CO$_2$ (3.3$\sigma$) and SO$_2$ (3.6$\sigma$) and a tentative detection of CH$_4$ (2$\sigma$). We calculate the detection significances by calculating the difference in $\chi^{2}$ from the best-fit model and a model where a particular molecule is removed. In Table\,\ref{tab:detection significances} we show the $\chi^{2}$ values. The lower panel shows the residuals with respect to the best-fit model. Scaled up error bars have been used to calculate the residuals. The solid and dashed lines show 1 and 2$\sigma$ confidence level. }
    \label{fig:full spectrum}
\end{figure*}

In Figure\,\ref{fig:full spectrum} we show the combined transmission spectrum of TOI-1130b. We have combined the JWST NIRSpec G395H and NIRISS SOSS transmission spectra (Section\,\ref{JWST data light curve analysis}). We also add the TESS and CHEOPS transit depths (Section\,\ref{tess and cheops}) for the complete optical/near-IR spectrum. The spectrum shows several features: we see absorption bumps around 1.4, 1.8 and 3\,$\mu$m that coincide with water absorption bands (marked with vertical shaded regions in Figure\,\ref{fig:full spectrum}). Along with this, we find absorption bumps around 4\,$\mu$m and 4.3\,$\mu$m that are co-incident with SO$_2$ and CO$_2$ absorption features respectively.  In the optical we find a large slope with NIRISS SOSS. The transit depths retrieved in SOSS are consistent with photometric TESS and CHEOPS transit depths.

\subsection{Free retrieval on TOI-1130b transmission spectrum} \label{free retrieval}

We run a free chemistry atmospheric retrieval with the publicly available code \texttt{PetitRadtrans} \citep{molliere2019} on the combined TESS+CHEOPS+JWST spectrum. We include H$_2$O, CH$_4$, CO$_2$, CO, SO$_2$, NH$_3$, HCN, H$_2$S and CS$_2$ mass fractions as free parameters in our forward model. These molecules were chosen as they have dominant absorption features in the wavelength range we probe and have been commonly included in atmospheric free retrievals of mini-Neptunes observed with JWST \citep[e.g. see][]{madhusudhan2023,benneke2024,beaty2024,rigby2025}. The mean molecular weight of the atmosphere is calculated from the mass fractions of the included species. We assume that the sum of molecular mass fraction equals 1. We assume that the remaining mass fraction is due to H and He and assume a solar H/He ratio.  We assume an isothermal atmosphere and include temperature as a free parameter. We fix the mass of the planet to 19.8~M$_{\oplus}$ \citep{borsato2024}. Since we see a strong slope in the optical part of the transmission spectrum, we add an aerosol opacity, which is modeled by a power-law:

\begin{equation}
    \kappa_\mathrm{cloud}=\kappa_{0}\,\lambda^{\alpha}
\end{equation}

where the coefficient of the opacity function ($\kappa_{0}$) and the power-law coefficient ($\alpha$) are included as fit parameters. We include an offset between the NIRISS and NIRSpec spectra \citep[e.g.\ see][]{madhusudhan2023}. Further details about the forward model are provided in Appendix\,\ref{appendix:retrievals}.

\begin{table*}
    \centering
    \begin{tabular}{c|c|c|c|c|c|c}
      Model   & $\delta\chi^{2}$ (no jitter) & $\delta\chi^{2}$ (with jitter) & $\sigma$ (no jitter) & $\sigma$ (with jitter) & lnB (Savage-Dickey) & $\sigma$ (Savage-Dickey)\\
      \hline\hline
      No $\mathrm{H_2O}$   &   91 & 56 & 9.5 & 7.5 & 22.1 & 6.9 \\
      No $\mathrm{CH_4}$   &  14 & 7.2 & 3.7 & 2.6 & 2.2 & 2.6\\
      No  $\mathrm{CO_2}$  &  17 & 10.7 & 4.2 & 3.3 & 3.8 & 3.2\\
       No $\mathrm{SO_2}$  &  15 & 14 & 3.9 & 3.6 & 5.3 & 3.6\\
    \end{tabular}
    \caption{Table showing the $\delta\chi^{2}$ and detection significance ($\sigma$) from our free retrieval with and without an added jitter term to account for residual red noise in our transmission spectrum. The $\delta\chi^{2}$ are calculated with respect to the best-fit model from the free retrieval and subsequently fixing the abundance of a particular molecule from the model to the lower limit of the uniform priors (10$^{-10}$). The approximate the Bayes factors are calculated using the Savage-Dickey approximation on the posterior distributions. The approximate Bayes factors are converted to a detection significance using the formalism outlined in \citet{trotta2008}. }
    \label{tab:detection significances}
\end{table*}

From the free retrievals we detect H$_2$O (9.5$\sigma$), CH$_4$ (3.7$\sigma$), CO$_2$ (4.2$\sigma$), SO$_2$ (3.9$\sigma$). We describe the calculation of detection significances in Appendix\,\ref{appendix:free retrieval}. The posterior distribution for CO, HCN, H$_2$S and CS$_2$ does not yield any constraints for these molecules. Although, other molecules, particularly organic molecules have been included in the retrieval of mini-Neptune JWST spectra \citep{rigby2025,constantinou2026}, we note that the precision of the transmission spectrum in this paper does not motivate inclusion of further molecular species with weaker features in the forward model. The posterior distribution from the free retrieval is shown in  Figure\,\ref{fig:free posterior}. This retrieval does not constrain the isothermal temperature and provides a 3$\sigma$ upper limit at 600~K. It has often been found that retrieved isothermal temperatures are lower than expected equilibrium temperature \citep[e.g.\ see][]{mayo2025,barat2025}, which could be due to averaging of morning and evening terminators \citep{MacDonaldryan2020}. Furthermore, in a free retrieval the only effect temperature has is to set the size of molecular features through the scale height \citep{de_wit}. In our models, both the radius at the reference pressure and isothermal temperature are free parameters. However, both can affect the size of molecular features and are therefore correlated, which can be seen in posterior distributions from this retrieval. This could be the reason why we do not constrain the isothermal temperature in the free retrieval, rather we find an upper limit. The free retrieval also finds a strong scattering slope with a 3$\sigma$ upper limit for the opacity power-law at -4. The offset between NIRSpec and NIRISS is consistent with 0. From this fit we calculate a mean molecular weight of 3.5. It is important to note that in the free retrieval the H$_2$O and CO$_2$ abundance pushed towards their upper bounds. Therefore, the mean molecular weight we calculate from this free retrieval can be interpreted as a lower limit. The retrieved parameters from this fit are provided in Table\,\ref{tab:table 1}. We find a reduced $\chi^{2}$ of 1.81.

Since the reduced $\chi^{2}$ is higher than 1, we redo the free retrieval by including an error inflation (`jitter') term in the likelihood function of our MCMC. Including this term relatively increases the width of the retrieved abundances. We detect H$_2$O (7.5$\sigma$), CO$_2$ (3.3$\sigma$), SO$_2$ (3.6$\sigma$) at comparatively lower significance compared to the fit without error inflation. The reduced $\chi^{2}$ from this fit is 1.01. We also have a tentative detection of methane (2.6$\sigma$). However, upon further investigation we found that the detection significance of CH$_4$ was being increased due to one outlier point at 3.8\,$\mu$m. When we remove that point from the retrieval the detection significance of CH$_4$ is reduced to 2$\sigma$, while the other molecular abundances and detection significances do not change significantly. Therefore, we adopt the molecular detection significances from this fit: H$_2$O (7.5$\sigma$),  CO$_2$ (3.3$\sigma$), SO$_2$ (3.6$\sigma$) and a tentative detection of CH$_4$ (2$\sigma$). We verify the detection significances by estimating the Bayes factor by post-processing the posterior distribution using the Savage-Dickey approximation \citep{dickey1971}, and using them to calculate detection significances using the formalism outlined in \citet{trotta2008}. We find that the detection significances are consistent with the calculation based on the difference in $\chi^{2}$.

We find that the red end of the NIRISS detector shows a large scatter and these data points appear as outliers. We test their impact on the retrieval results. We remove data points from 2.3--2.7\,$\mu$m and re-run the free retrieval using the same setup. We do not include a jitter term for these `test retrievals'. The posterior distribution for this is shown in Figure\,\ref{fig:free_posterior_outlier_removed} and the best-fit values are shown in Table\,\ref{tab:table 1}. We find that all the parameters are consistent within 1$\sigma$ compared to the retrieval on the full dataset. We find a reduced $\chi^{2}$ of 1.5, which is improved compared to the full dataset fit (without jitter). However, we find that the outliers at the red end of the NIRISS SOSS detector do not significantly affect the inferred abundances from the observed spectra.

We note that the free retrieval does not constrain the isothermal temperature of the planet, as it pushes down to very low values ($\sim$300~K). To test the effect of this, we fix the isothermal temperature to the expected equilibrium temperature (825~K) assuming 0 bond albedo and full recirculation and run the free retrieval. The posterior distribution and retrieved parameters are shown in Figure\,\ref{fig:free_ret_T_fixed} and Table\,\ref{tab:table 1} respectively. We detect all the molecules that were detected using the full retrieval, however, we find slightly higher abundance for CH$_4$ and SO$_2$ ($\sim$2$\sigma$ level of significance). We find a higher offset (25\,ppm) between NIRISS and NIRSpec compared to previous free retrievals. The water and CO$_2$ are consistent within 1$\sigma$ confidence level. We note that the reduced $\chi^{2}$ is 2.36, significantly higher than the full free retrieval. We conclude that the overall conclusion of a metal-rich atmosphere with detections of H$_2$O, SO$_2$ and CO$_2$ remains consistent even if we fix the isothermal temperature. 

We perform an independent atmospheric retrieval with free chemistry using the \texttt{PLATON} package \citep{Zhang:2024} on the combined JWST data only (without the TESS and CHEOPS data points), including the same combinations of molecular species as our \texttt{PetitRadtrans} analysis. We generate the opacity sources at a forward model resolution of R $= 20 000$. We include an opaque gray cloud deck and power-law haze as our aerosol treatment. The retrieved H$_2$O, SO$_2$ and CO$_2$ abundances are consistent with previous retrievals, while CO, CH$_4$ and NH$_3$ are unconstrained. The retrieved opaque gray cloud deck is significantly deeper than 10$^3$ bar, showing our previous runs with only a power-law opacity source is justified.  


From the free retrievals we can conclude that we detect three molecules (H$_2$O, CO$_2$, SO$_2$) at detection significance higher than 3$\sigma$ confidence level for all the free retrieval setups and a tentative detection of CH$_4$ (2$\sigma$). We find a strong slope in the optical, which can be explained by a power-law model. Our models find an upper limit to the isothermal temperature ($<$600~K at 3 $\sigma$), which is lower than the expected equilibrium temperature (825~K) assuming 0.1 albedo and full redistribution. The retrieved offset between NIRISS and NIRSpec observations is consistent with 0 at a 1$\sigma$ level of confidence. We post-process the posterior distributions and calculate mean molecular weight ($\mu$). Assuming the median values of the molecular abundances we calculate a $\mu$ of $\sim$3. However, we note that the posterior distributions for water and CO$_2$ asymptotically increase towards higher values of $\mu$, and we do not constrain the CO abundance. Therefore, we interpret the measured $\mu$ as a lower limit.



\begin{table*}
    \centering
    \begin{tabular}{c|c|c|c|c|c}
    \hline
    \hline
        Parameter name & All data & 2.3-2.7 removed & Platon & $T_\mathrm{eq}=825$~K & with jitter\\
        \hline
        \hline
        $\chi^{2}_\nu$ &   1.8 & 1.6 & 1.6 &  2.36 & 1.01 \\
       $[\mathrm{H_2O}]$  & -0.6$^{+0.3}_{-0.5}$ & -0.6$^{+0.2}_{-0.5}$ & -0.3$^{+0.2}_{-0.5}$& -0.4$^{+0.1}_{-0.2}$ & -0.7$^{+0.3}_{-1.1}$\\
      $[\mathrm{CH_4}]$    & -3.5$_{-0.7}^{+0.5}$ & -3.6$_{-1.1}^{+0.6}$ & $<-3.5$ & -2.5$_{-0.5}^{+0.4}$ & -4$_{-2}^{+1}$\\
       $[\mathrm{CO}]$              & $<-2$ & $<2$ & $<-0.6$ & $<-0.6$ & $<-2$\\
    $[\mathrm{CO_2}]$            & $-1.2^{+0.6}_{-0.9}$ & $-1.1^{+0.5}_{-0.8}$ & $-1.8^{+0.7}_{-0.8}$ & $-0.7^{+0.2}_{-0.5}$ & $-1.2^{+0.6}_{-1.9}$\\
    $[\mathrm{SO_2}]$            & $-3.0^{+0.6}_{-0.7}$ & $-3.0^{+0.6}_{-0.8}$ &-4.0 $^{+1.3}_{-1.5}$ & $-1.8^{+0.5}_{-0.7}$ & $-3.0^{+2.0}_{-1.1}$\\
    $[\mathrm{NH_3}]$            & $<-2$ & $<-2$ & $<-2$& $<-2$ & $<-2$\\
        $R$ [R$_{\oplus}$] & 3.65$^{+0.01}_{-0.01}$ & 3.64$^{+0.01}_{-0.01}$ & $3.57^{+0.01}_{-0.01}$ & 3.64$^{+0.01}_{-0.01}$ & 3.65$^{+0.02}_{-0.02}$\\
        $T_\mathrm{iso}$ [K] & $<$600 & $<$600 &$<$600 & 825 (fixed) & $<$600\\
       $\log{\kappa_\mathrm{cld}}$  & 2.3$^{+0.7}_{-0.9}$ & 2.5$^{+0.8}_{-1.1}$ &3.0$^{+0.6}_{-0.6}$ & 3.3$^{+0.5}_{-0.8}$ & 1.4$^{+1.0}_{-1.5}$\\
      $\alpha$   & $<$-4 & $<$-4 & $<$-4 & -9$_{-1.3}^{+2.4}$ & $<$3\\
      $P_\mathrm{cloud}$ &  - & - &5.4$_{-1.4}^{+1.3}$ & - & - \\ 
       Offset [ppm]  & -8$_{-7}^{+7}$ & -2$_{-9}^{+11}$ & $-14^{+7}_{-8}$ & -25$_{-9}^{+8}$ & -5$_{-12}^{+12}$\\
       $\mu$ [amu] & $>$3.5 &  $>$3.5 &$>$4.4 & $>$3.9 & $>$3\\

   \hline 
    \end{tabular}
    \caption{best-fit parameters from the free chemistry retrieval of TOI-1130b (See Section\,\ref{free retrieval}). The upper limits correspond to 3$\sigma$ upper limits. The abundances provided in this table are in mass fraction.  $\mu$ has been calculated from median values of detected molecular abundances and is interpreted as a lower limit. }
    \label{tab:table 1}
\end{table*}

\subsection{Equilibrium chemistry retrievals} \label{equilibrium chemistry retrieval}

We run an equilibrium chemistry retrieval using \texttt{PetitRadtrans}. All the molecules in the free retrieval were included. We note that the SO$_2$, which we detect with our free retrieval is produced due to photochemistry \citep{tsai2023}, and our equilibrium models do not produce significant amount of SO$_2$. However, self-consistent forward model grids have shown that H$_2$O and CO$_2$ are not expected to be strongly affected by photochemistry and vertical mixing for high metallicity atmospheres at the expected temperature of TOI-1130b \citep{mukherjee2024}. Furthermore, including disequilibrium chemistry in forward models can make them computationally expensive, which is not ideal for a retrieval. Therefore, taking these considerations into account we model the spectrum of TOI-1130b assuming equilibrium chemistry retrieval. In later sections we will discuss sparse grid models including self-consistent treatment of chemistry to validate the results we report in this section.

We assumed an isothermal atmosphere and include the temperature as a free parameter. We fixed the mass to 19.8~M$_{\oplus}$ \citep{borsato2024}. The radius at the reference pressure (0.01 bar) is kept as a free parameter. The chemistry is parameterized through the atmospheric metallicity and C/O ratio. Temperature, cloud deck opacity (power law model) and offset between NIRISS and NIRSpec are included as free parameters.

From this retrieval we constrain the atmospheric metallicity to $\log{Z}=1.92^{+0.5}_{-0.4}$ solar and the isothermal temperature to 645$^{+146}_{-50}$~K. The retrieved isothermal temperature is lower than the expected equilibrium temperature assuming a 0 Bond albedo (825~K) and full recirculation. The C/O ratio is constrained to a 3$\sigma$ upper limit of 0.75. The retrieved parameters are shown in Table\,\ref{tab:table 2}. We find a reduced $\chi^{2}$ of 1.9. Since the reduced $\chi^{2}$ is higher than 1, we re-run our equilibrium chemistry retrieval by including an error inflation term in our likelihood function, similar to the free retrieval case.

\begin{figure*}
    \centering
    \includegraphics[width=\linewidth]{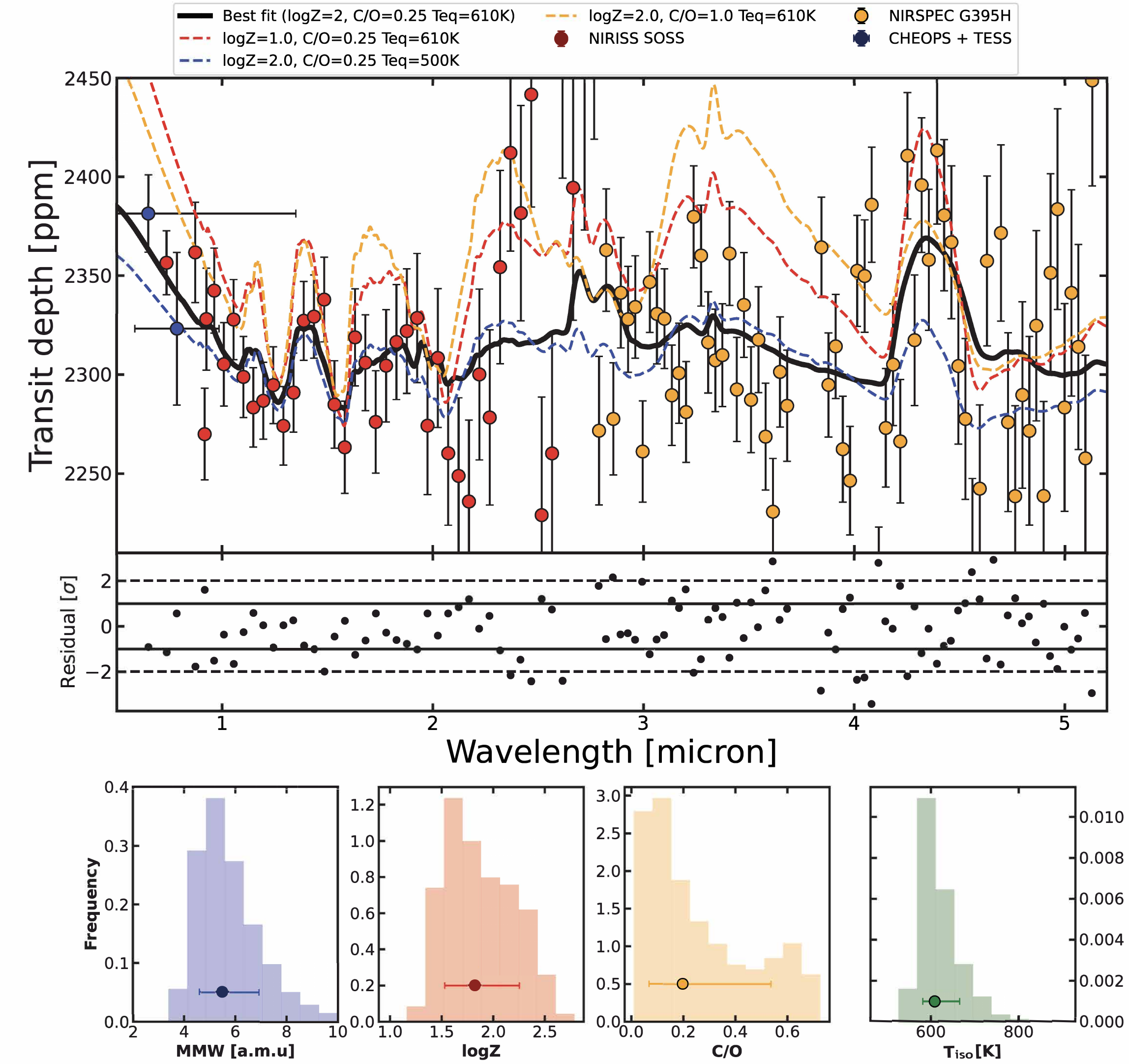}
    \caption{JWST/NIRSpec (orange) and NIRISS SOSS (red) transmission spectrum of TOI-1130b. The blue points show the transit depths derived from TESS and CHEOPS discussed in Section\,\ref{tess and cheops}. The derivation of JWST transmission spectra is discussed in Section\,\ref{JWST data light curve analysis}. The error bars shown for the observed spectra have been scaled up by the fitted jitter term from the equilibrium chemistry retrieval. The best-fit model has a ${\chi}^{2}_{\nu}$=1.1. The black solid model shows the best-fit model from an equilibrium chemistry retrieval implemented using \texttt{PetitRadtrans} \citep{molliere_19} including an error inflation term. The details of the forward model are outlined in Section\,\ref{equilibrium chemistry retrieval}. The different color models show the effect of varying independent parameters in our forward. We can rule out low atmospheric metallicity ($\sim$10), super-solar C/O ratio ($\sim$1), and very low isothermal temperature ($\sim$500~K) as they cannot match the CH$_4$ and CO$_2$ absorption features. The histograms in lower panels show the posterior distributions of key retrieved parameters. The mean molecular weight is derived by post-processing the metallicity and C/O posterior distributions. The best-fit parameters are shown in Table\,\ref{tab:table 2}.} 
    \label{fig:eq_chem_model}
\end{figure*}

In Figure\,\ref{fig:eq_chem_model} we show the best-fit model from the equilibrium chemistry retrieval with the error inflation term. The retrieved metallicity ($\log{Z/Z_\mathrm{\odot}}=1.8_{-0.3}^{+0.4}$) and C/O ratio ($0.2^{+0.4}_{-0.1}$) are consistent with the retrieval without error inflation and the values are shown in Table\,\ref{tab:table 2}. We adopt these values for the rest of the analysis. We show the retrieved posterior distribution for atmospheric composition and temperature in the same figure. We also show models with solar metallicity, super-solar C/O ratio (1.0) and lower isothermal temperature ($\sim$500~K). For  temperatures as low as 500~K, equilibrium models underpredicts the CO$_2$ absorption feature at 4.3\,$\mu$m and can be ruled out at 4$\sigma$ level of significance. Similarly, low metallicity ($\sim$10$\times$solar) atmospheres and super-solar C/O ratio models overpredicts CH$_4$ and underpredicts CO$_2$ absorption features and can be ruled out at $>$5$\sigma$ level of significance.


To test the effect of temperature we run two retrievals by fixing the isothermal temperature to 825~K (0 bond albedo) and 625~K (bond albedo 0.3). We find that for 825~K, our retrievals converge to a higher atmospheric metallicity (300$^{+200}_{-100}\times$solar) compared to a free temperature fit. The C/O ratio converges 0.6$\pm$0.02 ($\sim$solar). At higher temperatures, sub-solar C/O ratios ($\sim$0.2) do not produce methane. Therefore, to account for CH$_4$ the retrieval converges to a solar value. However, the reduced $\chi^{2}$ is 2.3 (higher than free isothermal temperature). For the 625~K retrieval, we find that the retrieved metallicity, C/O ratio and mean molecular weight are consistent with the free isothermal temperature fit. The reduced $\chi^{2}$ is 2 (slightly higher than the free temperature case).

From the equilibrium chemistry retrievals we conclude that TOI-1130b has a metal-rich atmosphere ($\log{Z/Z_{\odot}}=1.8_{-0.3}^{+0.4}$). We note that the posterior distribution for C/O ratio falls asymptotically, and we derive a 3$\sigma$ upper limit of 0.75. We post-process the retrieved posterior distribution for the equilibrium retrieval with the error inflation term and derived a mean molecular weight of 5.5$^{+1.3}_{-0.8}$ amu. We adopt the retrieved parameters from the equilibrium chemistry with the error inflation term for the rest of the analysis.



\begin{table*}
    \centering
    \begin{tabular}{c|c|c|c|c}
    \hline
    \hline
        Parameter name & Free $T_\mathrm{eq}$ & $T_\mathrm{eq}=825$\,K & $T_\mathrm{eq}=625$\,K & Free $T_\mathrm{eq}$ with jitter\\
        \hline
        \hline
       $\chi^{2}_{\nu}$ & 1.9 & 2.3& 2.0 & 1.1\\ 
       $\log{Z/Z_\odot}$ & 1.9$^{+0.5}_{-0.3}$ & 2.5$^{+0.2}_{-0.2}$ & 1.8$^{+0.3}_{-0.3}$ & 1.8$^{+0.4}_{-0.3}$\\
       C/O ratio & 0.2$^{+0.3}_{-0.1}$ & 0.6$^{+0.02}_{-0.02}$ & 0.2$^{+0.2}_{-0.1}$ & 0.2$^{+0.4}_{-0.1}$\\
       $R$ [R$_{\oplus}$] & 3.62$^{+0.01}_{-0.01}$ & 3.63$^{+0.007}_{-0.007}$ & 3.61$^{+0.007}_{-0.007}$ & 3.62$^{+0.007}_{-0.007}$\\
       T$_{eq}$ [K] & $645^{+146}_{-50}$ & 825 & 625 & $608^{+60}_{-30}$\\
        $\log{\kappa_\mathrm{cld}}$  & 3.1$^{+0.6}_{-0.9}$ & 2.5$^{+0.8}_{-1.1}$ & 3.0$^{+0.6}_{-0.8}$ & 2.9$^{+0.7}_{-1.0}$\\
      $\alpha$   &  -9.5$_{-1.4}^{+2.0}$ & -9.5$_{-1.8}^{+2.6}$ & -9.5$_{-1.5}^{+2.0}$ & -9.0$_{-1.7}^{+2.2}$ \\
       Offset [ppm]  & -10$_{-7}^{+7}$ & -17$_{-7}^{+7}$ & -13$_{-7}^{+7}$ & -13$_{-7}^{+8}$\\
       $\mu$ [amu] & 5.8$_{-1.1}^{+1.9}$ & 9.3$_{-2.6}^{2+.7}$ & 5.4$_{-1.1}^{+0.9}$ & 5.5$_{-0.8}^{+1.3}$\\
  \hline 
    \end{tabular}
    \caption{best-fit parameters from the equilibrium chemistry retrieval of TOI-1130b (See Section\,\ref{equilibrium chemistry retrieval}).}
    \label{tab:table 2}
\end{table*}

\subsection{Grid retrievals} \label{picaso grid}

We model the transmission spectrum of TOI-1130b using a self-consistent forward model grid generated with the 1D radiative–convective model \textsc{PICASO} \citep{2019Batalha_Picaso,2021Marley_Picaso,2023Mukherjee_Picaso} and the chemical kinetics model \textsc{Photochem} \citep{wogan23, wogan2024} including photochemistry. We generate temperature–pressure profiles and transmission spectra for TOI-1130\,b using this setup. 

We ran a grid of 27 models testing metallicities of \(1\times\), \(10\times\), and \(100\times\) solar, C/O ratios of 0.3, 1.0, and 2.0, and \(K_{zz}\) values of \(10^6\), \(10^{10}\), and \(10^{13}\,\mathrm{cm^2\,s^{-1}}\). Within each model, \textsc{PICASO} calculated the temperature structure and radiative transfer, and \textsc{Photochem} provided disequilibrium abundances. We first compute a PT structure with \textsc{PICASO} for each grid model using the system’s stellar and planet parameters. Stellar and planetary parameters were taken from \citet{borsato2024} -\(T_\mathrm{s}=4360~\mathrm{K}\), \(\log {g}=4.55\), \([\mathrm{Fe/H}]=+0.11\), \(R_\mathrm{s}=0.697~R_\odot\), and \(M_\mathrm{p}=19.8~M_{\oplus}\). We set \(T_\mathrm{eq}=825~\mathrm{K}\) and \(T_\mathrm{int}=50~\mathrm{K}\). To match the infrared transit depth, we adopt \(R_\mathrm{p}=3.55~R_{\oplus}\).

Each atmosphere used 91 plane-parallel layers, logarithmically spaced in pressure between \(10^{5}~\mathrm{bar}\) and \(10^{-6}~\mathrm{bar}\). We then ran \textsc{Photochem} on the same vertical grid with \(K_{zz}\) to include kinetics and photochemistry. We used a MUSCLES UV and optical spectrum for the nearby K-dwarf HD~85512. This is not the exact TOI-1130 SED, but it provides realistic UV for photochemistry and haze precursors. We rescaled the MUSCLES spectrum so that the flux at the planet corresponds to \(T_{\mathrm{eq}} = 825~\mathrm{K}\), assuming zero albedo and full redistribution. This way we can model the UV and optical input to the atmosphere consistently. The resulting disequilibrium mixing ratios were passed back to \textsc{PICASO} for the transmission spectrum calculation.

We compare the 27 self-consistent forward model models with the observed transmission spectrum and find that the best-fit model has 100$\times$solar metallicity, C/O ratio of 0.3 and K$_{zz}$=10$^{10}$cm$^2$/s. The metallicity and C/O ratio values are consistent with the equilibrium chemistry retrievals from \texttt{PetitRadtrans} (Section\,\ref{equilibrium chemistry retrieval}). We find a reduced $\chi^{2}$=3.68. However, the baseline models did not reproduce the very steep short-wavelength slope. To address this, we tested some modifications, that were applied individually and in combinations, while keeping the PT structure and baseline composition fixed.



First, within the coupled \textsc{PICASO} and \textsc{Photochem} model, we examined the abundances of the sulfur species, including \(\mathrm{SO_2}\) observed at \(4\,\mu\mathrm{m}\).
Photochemical reactions create layers dominated by \(\mathrm{SO_2}\) at \(10^{-6}\) to \(10^{-4}~\mathrm{bar}\), \(\mathrm{S_2}\) at \(10^{-4}\) to \(10^{-2}~\mathrm{bar}\), with \(\mathrm{H_2S}\) dominating at deeper layers.
The layer we are probing in the transmission spectrum near \(4\,\mu\mathrm{m}\) is about \(0.06\) to \(0.10~\mathrm{mbar}\).
At that layer, \(\mathrm{H_2S}\) is the dominant sulfur carrier.
To fit the observed \(\mathrm{SO_2}\) abundance, we increase the sulfur inventory by \(\sim 10\times\) by multiplying all S-bearing species by 10 and renormalizing each layer so the mixing ratios sum to one. The boost raises \(\mathrm{SO_2}\) and improves the fit near \(4\,\mu\mathrm{m}\), and we use the boosted spectrum in the model comparison. In the best-fit model, \(\mathrm{SO_2}\) reaches a volume mixing ratio of \(5.17\times10^{-5}\) at the pressures probed by the \(4\,\mu\mathrm{m}\) feature.

SO$_2$ abundance depends strongly on atmospheric metallicity \citep{mukherjee2024}. At temperatures around 800~K SO$_2$ production is inefficient at solar metallicity \citep{crossfield2025}. Therefore, the detection of SO$_2$ in the atmosphere of TOI-1130b independently shows that it has a high metallicity atmosphere.

\begin{figure}
    \centering
    \includegraphics[width=1\linewidth]{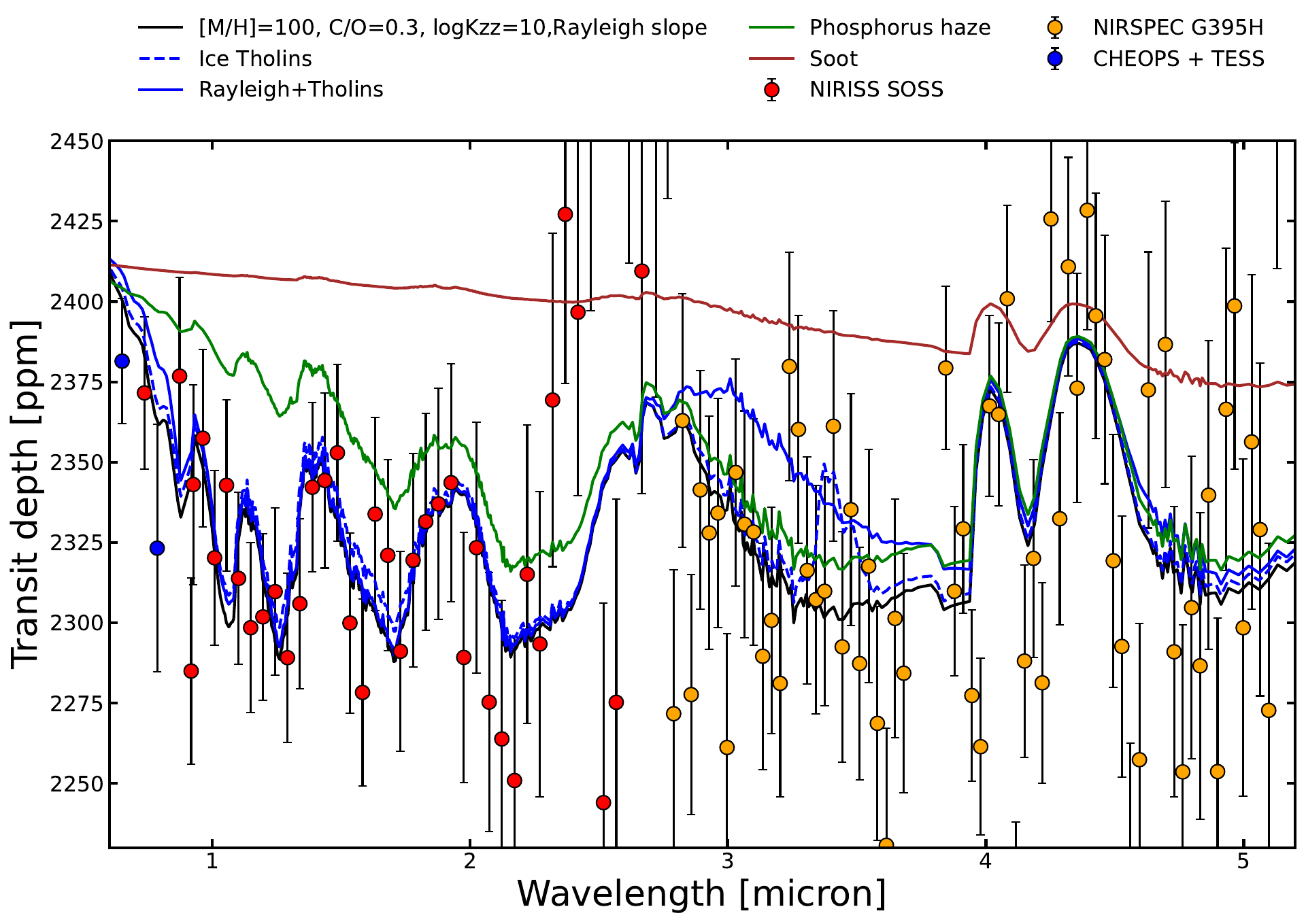}
    \caption{The transmission spectrum of TOI-1130b observed with JWST NIRISS and NIRSpec with best-fit models (100$\times$solar, 0.3 C/O ratio and K$_{zz}$=10$^{10}$cm$^2$/gm) from the PICASO grid (Section\,\ref{picaso grid})
    . A Rayleigh slope and Sulfur boost has been added to fit the observations. The best model reduced $\chi^{2}$ is 2.53. We also different haze models: Titan-like tholins \citep{khare1984}, comet-like ice-tholins \citep{khare1993}, soots \citep{lavvas2017} and Saturn-like Phosphorus haze \citep{noy1981}. A comparison between these models is shown in Table\,\ref{tab:haze models}}
    \label{haze models}
\end{figure}

We also added a Rayleigh-like power law in the upper atmosphere to match the short-wavelength optical slope,
\[
\tau_{\mathrm{Rayleigh}}(\lambda)=\tau_0\left(\frac{\lambda}{\lambda_0}\right)^{-\alpha},
\]
with \(\tau_0=0.04\) at \(\lambda_0=0.5\,\mu\mathrm{m}\) and \(\alpha=4.5\). We applied this to the top 20 layers setting the single scattering albedo to \(\omega_0\approx0.999\) and the asymmetry parameter to \(g_0\approx0\) to add a steep, mostly isotropic scattering slope that matches the observed short wavelength rise in transit depth. \textsc{PICASO} adds this as an extra scattering opacity in the transmission calculation. With these modifications we could model the optical slope and fit the SO$_2$ feature better compared to the baseline model. We improve the residual $\chi^{2}$ to 2.53. This residual $\chi^{2}$ is comparable to the free chemistry and equilibrium chemistry fits with a fixed (825~K) isothermal temperature. The best-fit model from our grid with  scattering and boosted Sulphur abundance is shown in Figure\,\ref{haze models}. We note that the power-law index used for these models is consistent with the upper limit found from the atmospheric retrievals. We also tested models with $\alpha$=-4.25 and -4 and we found that the $\chi^{2}_{\nu}$ did not change significantly.

We also tested different haze slabs with \textsc{Virga} \citep{Virga}. We used a lab size distribution with median radius \(\sim26\,\mathrm{nm}\) and a 15 to 60\,nm span \citep{He2018a}. For each haze sample, \textsc{Virga} used the haze specific refractive indices and used \texttt{PyMieScatt} \citep{sumlin2013} to compute wavelength-dependent Mie efficiencies and the asymmetry parameter. \textsc{PICASO} mapped these inputs to per-layer \(\tau_\mathrm{ext}\), \(\omega_0\), and \(g_0\) on its cloud grid and computed limb-viewing transmission spectra. We have tested different haze compositions, such as Titan-like reflective tholin hazes \citep{khare1984}, which produce a strong feature around 3.3\,$\mu$m and soots \citep{lavvas2017}, which do not reproduce molecular absorption features, Saturn-like Phosphorus hazes \citep{noy1981} and ice-tholins, found on icy bodies of our solar system \citep{khare1993}. In Table\,\ref{haze models} and Figure\,\ref{haze models} we compare these different haze models. We find that we can rule out tholin, soot and Phosphorus hazes with high statistical significance($>$5$\sigma$ level of confidence). We also find that a ice-tholins model is worse than base-line power-law model at 3$\sigma$.

\begin{table}
    \centering

    \begin{tabular}{c|c|c}
        \hline
        \hline
       Model  & $\chi^{2}$ & Rejection $\sigma$=$\sqrt{\Delta\chi^{2}}$ \\
       \hline
      Baseline+Rayleigh   & 253 &  - \\
       Ice-Tholin  & 261 & 3.09  \\
       Tholin  & 325 &  8.55 \\
       Soot  & 1199 &  30.90 \\
        Phosphorus & 435 &  13.53 \\
        \hline
    \end{tabular}
    \caption{Comparing the different haze models with the observed spectrum of TOI-1130b. The $\Delta\chi^{2}$ values are computed with respect to the baseline model, which is used to calculate the rejection significance of the different models. The number of degrees of freedom for each model are same. }
    \label{tab:haze models}
\end{table}

\subsection{Comparison between model fits to TOI-1130b transmission spectrum} \label{model_comparison}

\begin{figure*}
    \centering
    \includegraphics[width=1\textwidth]{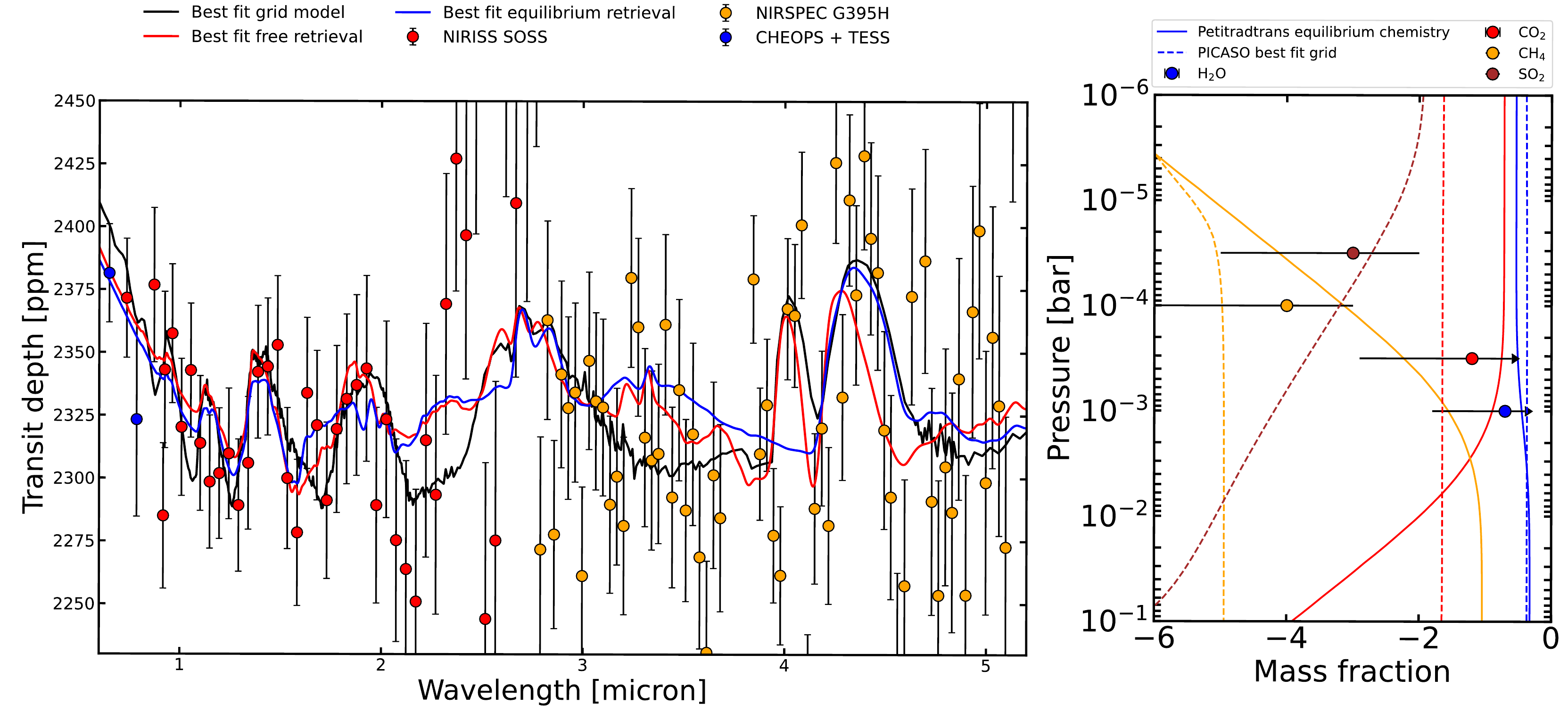}
    \caption{Transmission spectrum of TOI-1130b with the best-fit free chemistry, equilibrium chemistry and grid model overplotted  (left) and comparison between retrieved abundances from the three model fits for TOI-1130b (right). The points shown on the right panel are the best fit mass fractions from the free retrieval. The H$_2$O and CO$_2$ posterior distributions shows a lower limit, where as the other molecules show gaussian distributions. Therefore H$_2$O and CO$_2$ data points are shown with upper arrows. The solid lines show chemical abundance profiles from equilibrium chemistry and the dashed lines show the chemical profiles from the self-consistent forward model grid.}
    \label{model comparison}
\end{figure*}

We have modeled the combined TESS CHEOPS and JWST transmission spectrum TOI-1130b using a free chemistry, equilibrium chemistry retrieval and a self-consistent forward model grid. We compare the three models in Figure\,\ref{model comparison}. We find that the retrieved free chemistry abundances are consistent with both the equilibrium chemistry and self consistent forward model grid abundance profiles. We find H$_2$O (7.5$\sigma$),  CO$_2$ (3.3$\sigma$), SO$_2$ (3.6$\sigma$) and a tentative detection of CH$_4$ (2$\sigma$). The free retrieval model and the grid model (including photochemistry) show the SO$_2$ feature at 4\,$\mu$m. Since SO$_2$ is mainly produced due to disequilibrium photochemistry, our equilibrium chemistry model does not produce observable amount of SO$_2$. The retrieval models have been fitted using an error inflation term. The metallicity and C/O ratio we derive from the equilibrium chemistry model and grid model are consistent with each other ($\sim$100$\times$solar, C/O $\sim$ 3$\sigma$ upper limit of 0.75). This results in a $\mu$ of 5.5$^{+1.3}_{-0.8}$ amu. This is also consistent with the free retrieval that finds a lower limit of 3 amu.  For the free retrieval, the isothermal temperature is low ($\sim$400~K) compared to the equilibrium chemistry retrieval and grid model. Therefore, we find that the width of the 4.3\,$\mu$m CO$_2$ feature is smaller in the free retrieval best-fit model. We conclude that considering the differences in all three fitting approaches, all three agree to a metal-rich high mean molecular weight atmosphere.

\subsection{Potential He escape from TOI-1130b} \label{helium escape}

\begin{figure}[!htpb]
    \centering
    \includegraphics[width=\linewidth]{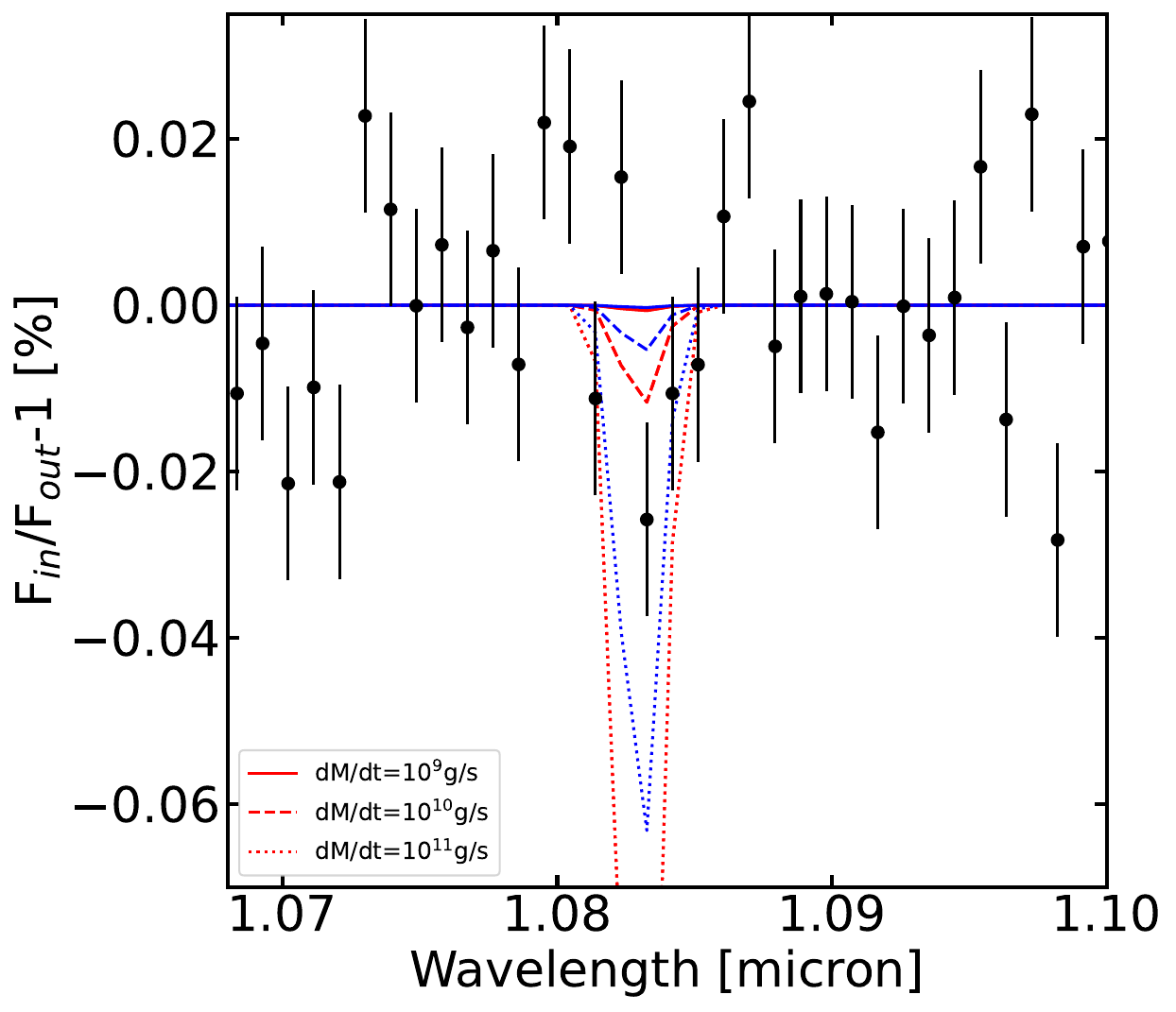}
    \caption{We show the transmission spectrum of TOI-1130b, zoomed in on the He triplet feature. We derive this spectrum using a formalism described in Section\,\ref{helium escape}. The overplotted models have been calculated using the publicly available p-winds \citep{does_santos2022} code. Red lines correspond to upper atmosphere temperature of 8000~K and blue lines correspond to 10000~K models. The different line styles correspond to different mass loss rates.}.
    \label{fig:He zoom in}
\end{figure}

In Figure\,\ref{fig:He zoom in} we show the transmission spectrum of TOI-1130b, zoomed in on the 1.083\,$\mu$m He triplet feature, which has been used as a tracer of atmospheric escape \citep{oklopcic2018,spake18}. We do not find a significant planetary He absorption feature. We use the formalism outlined in \citep{orell2024} to derive the transmission spectrum. We take a 20 pixel half-width region centered on the He line and normalize the stellar 1D spectrum by fitting a straight line. From the continuum normalized spectrum, we calculate the mean in-transit spectrum and mean out of transit spectrum. 

We compare this spectrum with models calculated using the publicly available \texttt{p-winds} code \citep{does_santos2022}. We adopt the stellar SED for HD 85512 from the MUSCLES survey \citep{france2016}. We calculate a grid of models assuming two different temperatures of the upper atmosphere (8000~K and 10000~K) and three mass loss rates (10$^{9}$, 10$^{10}$, 10$^{11}$ g/s). We find that for both temperatures high mass loss rates (10$^{11}$ g/s) produce a deep He absorption feature and can be ruled out by the observed spectrum.


\section{Discussion} \label{discussion}

\subsection{Origin of volatiles in TOI-1130b: accretion beyond water ice line} \label{formation_discussion}

The orbital architecture (inner mini-Neptune and outer hot Jupiter) and the 2:1 mean motion resonance of the TOI-1130 system provides strong hints that this system assembled beyond the water ice line and migrated inwards due to disk migration. It has been hypothesized that mini-Neptunes could have originated beyond the ice line and later migrate \citep{raymond2018,venturini2020,zeng2021}. In this case, the planet would have access to volatile-rich solids that can be accreted to increase the volatile budget of the planet \citep{bitsch2019,bitsch2021}. This would lead to volatile-rich interiors \citep{aguichinne2021} and atmospheres \citep{aguichine2025}. 

The high mean molecular weight atmosphere (5.5$^{+1.3}_{-0.8}$ amu) our transmission spectrum has revealed for TOI-1130b indicates that it is volatile-rich: 72\% H$_2$/He and 20\% H$_2$O by mass. This is consistent with formation beyond the ice line and significant accretion of volatile-rich solids and disk gas rich in H$_2$/He to form a high mean molecular weight envelope. The disk within the water ice line can be enriched with volatiles due to pebble drift and evaporation \citep{booth2017,schneider2021}, however due to the presence of the hot Jupiter in the TOI-1130 system, the migrating pebbles would be blocked \citep{bitsch2021}, leaving it volatile-poor. Therefore, if TOI-1130b accreted within the water ice line, it would be difficult to explain its observed atmospheric metal enrichment. Furthermore, it has been shown that in-situ formation results in low C/O ratio \citep{steinmeyer2026}, and the detection of both H$_2$O and CO$_2$ in TOI-1130b is consistent with formation beyond the ice line.

TOI-1130b occupies a rare location in the radius-period diagram: with a radius of 3.66~R$_{\oplus}$ and orbital period of $\sim$4 days, it lies at the edge of the radius cliff with an occurrence rate of $<$1\% \citep{Dattilo2023}. One of the leading theories for the formation of mini-Neptunes is the `gas-dwarf scenario' \citep{owen2013,owen2017}:  in-situ formation \citep{lee2014,lee_chiang2016} and subsequent evolution due to mass loss mechanisms \citep{owen_jackson2012,ginzburg2018}. These mechanisms naturally remove significant amount of H/He from close-in small planets and sculpt them to smaller radii, leading to the dearth of planets larger than 3~R$_{\oplus}$ and forming the radius cliff. However, recent studies have pointed out that mass loss processes alone cannot explain the radius cliff \citep{dattilo2024,lee2025} and may require different formation mechanisms, such as formation beyond the ice line and migration \citep{chakraborty2026,burn2024}. If, TOI-1130b was born with a high mean molecular weight, it would not undergo significant atmospheric mass loss \citep{yoshida2025} and would retain its size through its evolutionary phase \citep{rogers2025}. This could explain its position on the radius-period diagram. The non-detection of planetary He absorption (Figure\,\ref{fig:He zoom in}) is consistent with this. Assuming a mass loss rate upper limit of 10$^{11}$g/s, we calculate that the planet would lose $\sim$0.5M$_{\oplus}$ over 1 Gyr, and therefore not undergo significant evolution.


\subsection{Alternative source of volatiles in TOI-1130b} \label{outgassing_discussion}

Recent studies have shown that interior-atmosphere interactions could have a significant impact on the atmospheric composition of low mass planets \citep{lichtenberg2025,shorttle2024,rigby2024}, like mini-Neptunes, as their interior and envelope mass fractions are comparable.  For these mini-Neptunes, it has been suggested that global magma oceans can exist at the interior atmosphere boundary \citep{dorn2021}, and interactions between the magma and atmosphere can outgas significant amount of volatiles, like water \citep{kite2020}. These interactions can affect metal mass fraction \citep{werlen2025} and elemental abundance ratios \citep{werlen2025b,steinmeyer2026}, which have been traditionally linked to planet formation \citep{oberg_2011}. For instance, \citet{nixon2025} have shown that the volatiles detected in the mini-Neptune TOI-270d \citep{benneke2024} can be explained by magma-atmosphere interactions.

However, these models are well suited for low mass mini-Neptunes. TOI-1130b, with an estimated envelope mass fraction of 25\%, assuming 15~M$_{\oplus}$ core, and a high mean molecular weight atmosphere ($\sim$5.5) lies at a transition between low mass mini-Neptunes and more massive Neptune-like planets. It has been suggested that for planets with high pressure at the interior-envelope boundary, magma is likely to solidify \citep{breza2025,calder2025}, and cut-off interior atmosphere interactions. Comparing TOI-1130b with grids presented in \citet{breza2025}, we find that it lies in a region where it is difficult to predict whether it will have a liquid magma ocean. Furthermore, strong vertical mixing is necessary to affect the atmospheric composition of the observable atmosphere, even if magma-atmosphere interactions are happening \citep{werlen2025b}. Therefore, for planets with high envelope mass fraction, such as TOI-1130b, current models do not predict how strongly their observable atmospheres can be affected due to these interactions.

\subsection{Comparing TOI-1130b with mini-Neptunes/super-Neptunes} \label{sub neptune classification}

TOI-1130b lies at the edge of the radius cliff; a transition between the smaller mini-Neptunes and the larger and rarer super-Neptunes. In this section, we compare the inferred atmospheric properties of TOI-1130b with both mini-Neptunes and super-Neptunes.

\subsubsection{Mean molecular weight}

\begin{figure*}
    \centering
    \includegraphics[width=\linewidth]{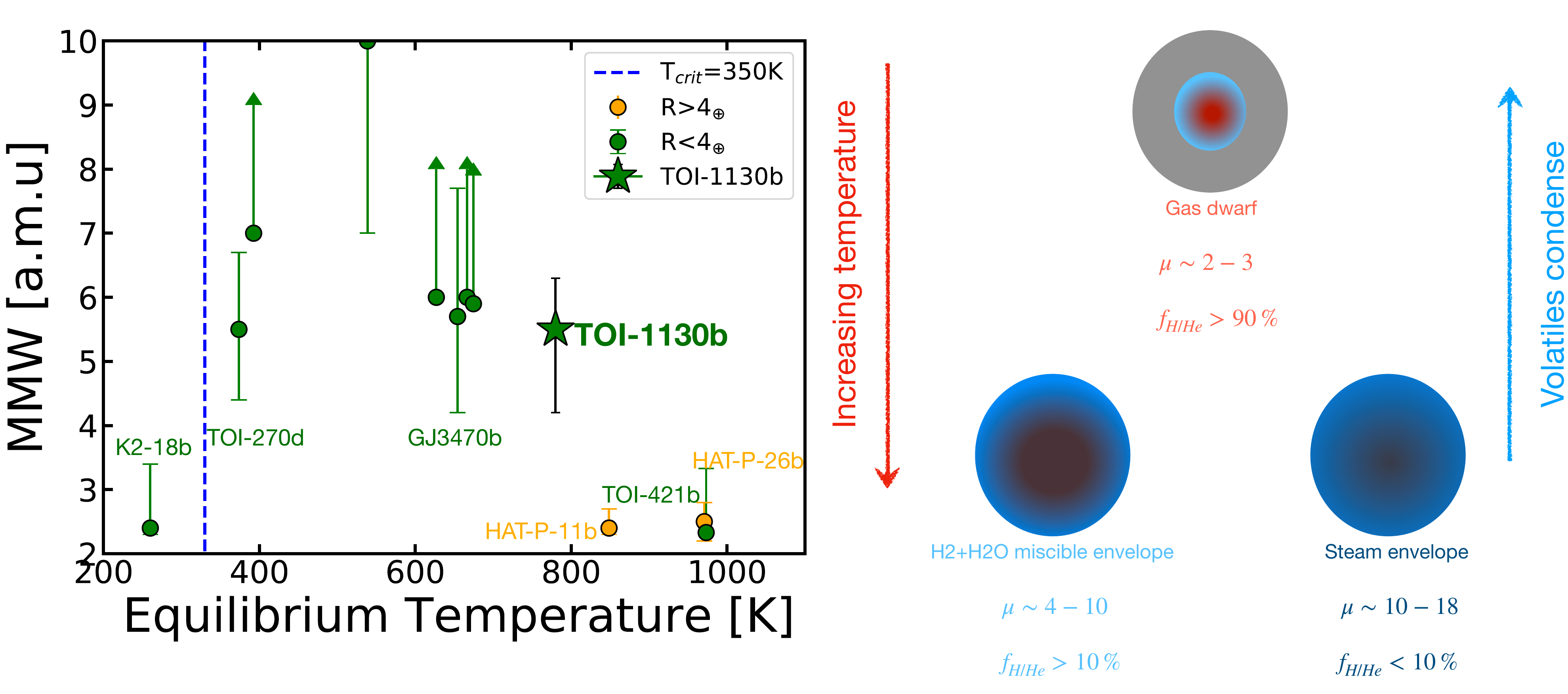}
    \caption{Mean molecular weight ($\mu$) of mini/super-Neptunes as a function of equilibrium temperature. The $\mu$ are taken from literature and shown in Table \ref{tab:MMW table}. We show TOI-1130b on this figure. We calculate the mean molecular weight from the posterior distribution of the equilibrium chemistry retrieval (Section\,\ref{equilibrium chemistry retrieval}). The blue vertical line marks the critical temperature for condensation of water (350~K) as proposed in \citet{madhusudhan2025b}. The schematic on the right demonstrates this idea. Above the critical temperature, the water in the planet is vapourized and mixed with the H/He envelope to form either H$_2$-H$_2$O mixed envelopes (miscible envelope mini-Neptune) or steam dominated envelopes (steam worlds). The intermediate $\mu$ of TOI-1130b likely aligns it with the miscible envelope mini-Neptunes.}
    \label{fig:mmw figure}
\end{figure*}

TOI-1130b has a mean molecular weight of 5.5$^{1.3}_{-0.8}$ derived from the best-fit retrieval models. In Figure\,\ref{fig:mmw figure} we compare its mean molecular weight with mini-Neptunes and super-Neptunes with constrained $\mu$ from HST/JWST observations. The sample of planets used for this comparison is shown in Table\,\ref{tab:MMW table}. TOI-1130b has similar $\mu$ to mini-Neptunes like TOI-270d ($\sim$5.5, \citealt{benneke2024}) and GJ3470b ($\sim$6, \citealt{beaty2024}). It has been hypothesized that the atmospheric mean molecular weight is dependent on the temperature \citet{benneke2024,madhusudhan2025b}. Above a critical temperature ($\sim$350~K), water is vapourized and mixes into the atmosphere with lighter gases like H$_2$ and He. These warm mini-Neptunes have been classified as `miscible envelope mini-Neptunes' \citep{benneke2024}. Although TOI-1130b is more massive than typical mini-Neptunes like TOI-270d, its atmospheric $\mu$ makes it consistent with the miscible envelope scenario. Compared to the mini-Neptunes, super-Neptunes like HAT-P-11b \citep{chachan_2019} and HAT-P-26b seem to have low $\mu$ atmospheres, comparable to H/He dominated atmospheres of gas giants. Therefore, although TOI-1130b is comparable to HAT-P-11b (25~M$_{\oplus}$) and HAT-P-26b (19~M$_{\oplus}$) in terms of mass, their atmospheres appear very different (volatile-rich versus H/He dominated). With more planets being observed near the radius cliff in future we will be able to test whether truly a transition in atmospheric composition occurs in this region.

\subsubsection{SO2 detection in mini-Neptunes} \label{sulpher abundance}

\begin{figure}
    \centering
    \includegraphics[width=1\linewidth]{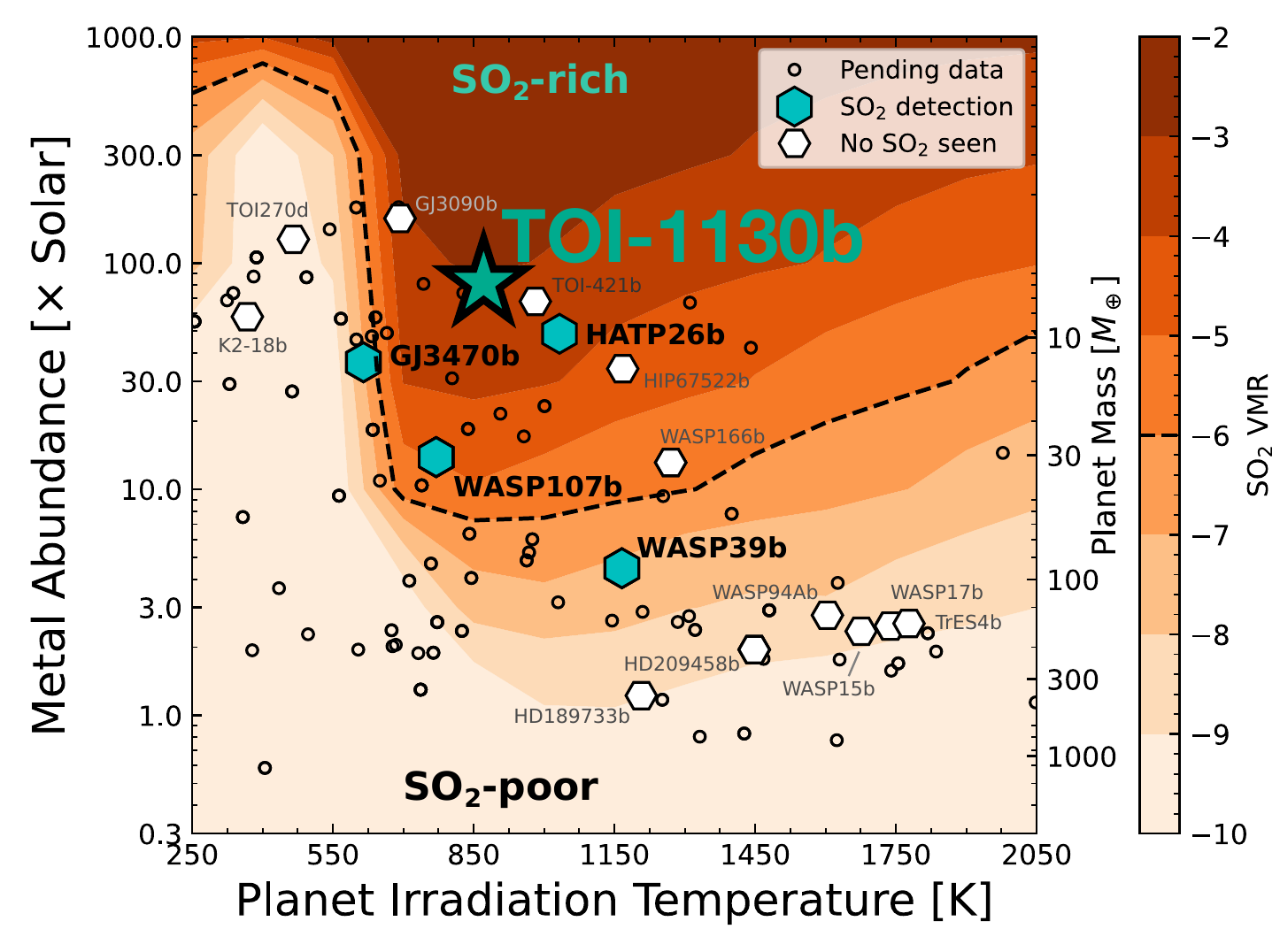}
    \caption{SO$_2$ abundance as a function of irradiation temperature and atmospheric metallicity (left) adapted from \citet{crossfield2025}. TOI-1130b joins GJ3470b as the second mini-Neptune with a SO$_2$ detection, and is consistent with a high metallicity atmosphere.}
    \label{fig:sulpher_figure}
\end{figure}

From the free retrievals we constrain the SO$_2$ abundance to a mass fraction of 10$^{-3}$ and a volume mixing ratio (VMR) of 10$^{-4}$. In Figure\,\ref{fig:sulpher_figure} we show TOI-1130b in the atmospheric metallicity-irradiation temperature plane with SO$_2$ abundance contours from self-consistent forward model grids that include sulphur photochemistry \citep{crossfield2023,crossfield2025}. TOI-1130b joins GJ3470b \citep{beaty2024} as a second mature mini-Neptune with a detected SO$_2$ feature. 

SO$_2$ production is strongly dependent on equilibrium temperature and atmospheric metallicity \citep{mukherjee2024}. For TOI-1130b, the predicted SO$_2$ abundance is very low for metallicities below 30$\times$solar. Therefore, the SO$_2$ detection independently rules out such low metallicity atmospheres, consistent with our finding from equilibrium chemistry retrievals. We also find that for temperatures below 600~K SO$_2$ production requires extremely high metallicity ($>$500$|times$solar). This is consistent with non detections reported for cooler mini-Neptunes, such as TOI-270d \citep{benneke2024} or K2-18b \citep{madhusudhan2023}, and detections for warmer min-Neptunes like GJ3470b and TOI-1130b.

\subsubsection{Atmospheric feature size as a function of temperature} \label{sub neptune haze}

A quadratic temperature dependent trend has been reported for molecular absorption feature size for mini-Neptunes/super-Neptunes \citep{Brande2024}. This trend has been reported based on the 1.4\,$\mu$m water absorption feature accessible to HST/WFC3. The fitted trend shows that the water feature amplitude size is attenuated between 500--700~K. 

We re-examine this trend using a sample of mini-Neptunes and super-Neptunes shown in Table\,\ref{tab:subneptune_final}. We calculate their expected equilibrium temperature assuming an albedo of 0.3 (as assumed in \citet{Brande2024}). For planets with published JWST transmission spectra, we assume the best-fit value or lower limit on $\mu$ from literature for the calculation. For those planets without a published mean molecular weight we assume 100$\times$solar metallicity and solar C/O ratio, which approximately translates to a $\mu$ of 4. We define the scale height metric in Eqn\,\ref{eq:eqn1} \citep{Brande2024},

\begin{equation*} \label{eq:eqn1}
    A_{H}\equiv(\delta_{\lambda_{0}}-\delta_{\lambda_{1}})(R^{2}_{*}/2\,H\,R_{p}) \ .
\end{equation*}
We calculate $A_{H}$ for the 1.4\,$\mu$m water band ($\lambda_{0}=1.4\,\mu$m, $\lambda_{1}=1.25\,\mu$m), and for the CO$_2$ band ($\lambda_{0}$=4.3, $\lambda_{1}$=4.15). $\delta_{\lambda}$ corresponds to the transit depth at a particular wavelength. The wavelengths in our calculation are chosen such that we compare the center and wings of a molecular feature. 
For TOI-1130b we calculate an equilibrium temperature of 780~K (assuming 0.3 albedo and full recirculation) and a $\mu$ of 5.5 for this calculation. We find $A_{H}^\mathrm{H_2O}$=2.3$\pm$1.3 and $A_{H}^\mathrm{CO_2}$=6.4$\pm$2.1 for TOI-1130b. The values and data used for calculation of $A_H$ for the other planets in the sample is shown in Table\,\ref{tab:subneptune_final}.

 We re-derive the quadratic trend by including new JWST observations that had not been used in \citet{Brande2024}. We find that if we include data points below 1000~K, the new fit parameters are consistent with \citet{Brande2024} fit within 2$\sigma$, however, if we include planets hotter than 1000~K, we find a much weaker quadratic trend. The quadratic term is consistent with 0. TOI-1130b is consistent with the quadratic $A_{H}^\mathrm{H_2O}$ trend presented in \citet{Brande2024} as well as the new fit we derive. We do not find a significant trend for $A_{H}^\mathrm{CO_2}$. We find that a flat line fit for $A_{H}^\mathrm{CO_2}$ gives $\chi^{2}_{\nu}$=1.07 whereas a quadratic fit gives 1.4. The $A{_H}^\mathrm{H_2O}$ trend seems consistent with both mini-Neptunes and super-Neptunes cooler than 1000~K.

One of the potential explanations for the quadratic $A_{H}^\mathrm{H_2O}$ trend is the presence of photochemical hazes \citep{gao2020b}. Haze production becomes efficient at temperatures around 600~K due to increasing methane abundance\citep{gao2020,fortney_2020} and decreases with increasing temperature \citep{He2018b}. At low temperatures ($<400$\,K) lab experiments have shown that haze production could decrease \citep{yu2021}. Haze opacity decreases at longer IR wavelengths \citep{kawashima2018}. Therefore, the $4.3\,\mu$m CO$_2$ feature could be less attenuated compared to the $1.4\,\mu$m water feature, which is consistent with the non-detection of a trend for $A_{H}^\mathrm{CO_2}$. It has been predicted that for equilibrium temperatures higher than 950\,K the aerosol opacity is dominated by silicate (SiO$_2$) condensation \citep{gao2020b}. Silicate clouds have been detected for hot Jupiters\citep{grant2023,inglis2024} and directly imaged planets \citep{miles2023,hoch2025} in the temperature range of 1000-1500\,K. We find that including planets with temperature higher than 1000~K weakens the quadratic trend in the water band. Interestingly, the only data point in our sample that is ~1000~K and has a CO$_2$ feature also lies below the best-fit trend. This could potentially be due to the formation of silicate clouds at temperatures around 1000~K, which would have an effect on both bands discussed here. Therefore, at temperatures around 1000~K we could see a transition between the dominant aerosol composition from hazes to clouds, and clouds would attenuate the molecular features across all wavelengths, unlike hazes that affect blue wavelengths more. However, we note that the sample size of planets are small and we need future JWST observations in this parameter space to confirm this potential trend.

\begin{figure}[!htpb]
    \centering
    \includegraphics[width=1\linewidth]{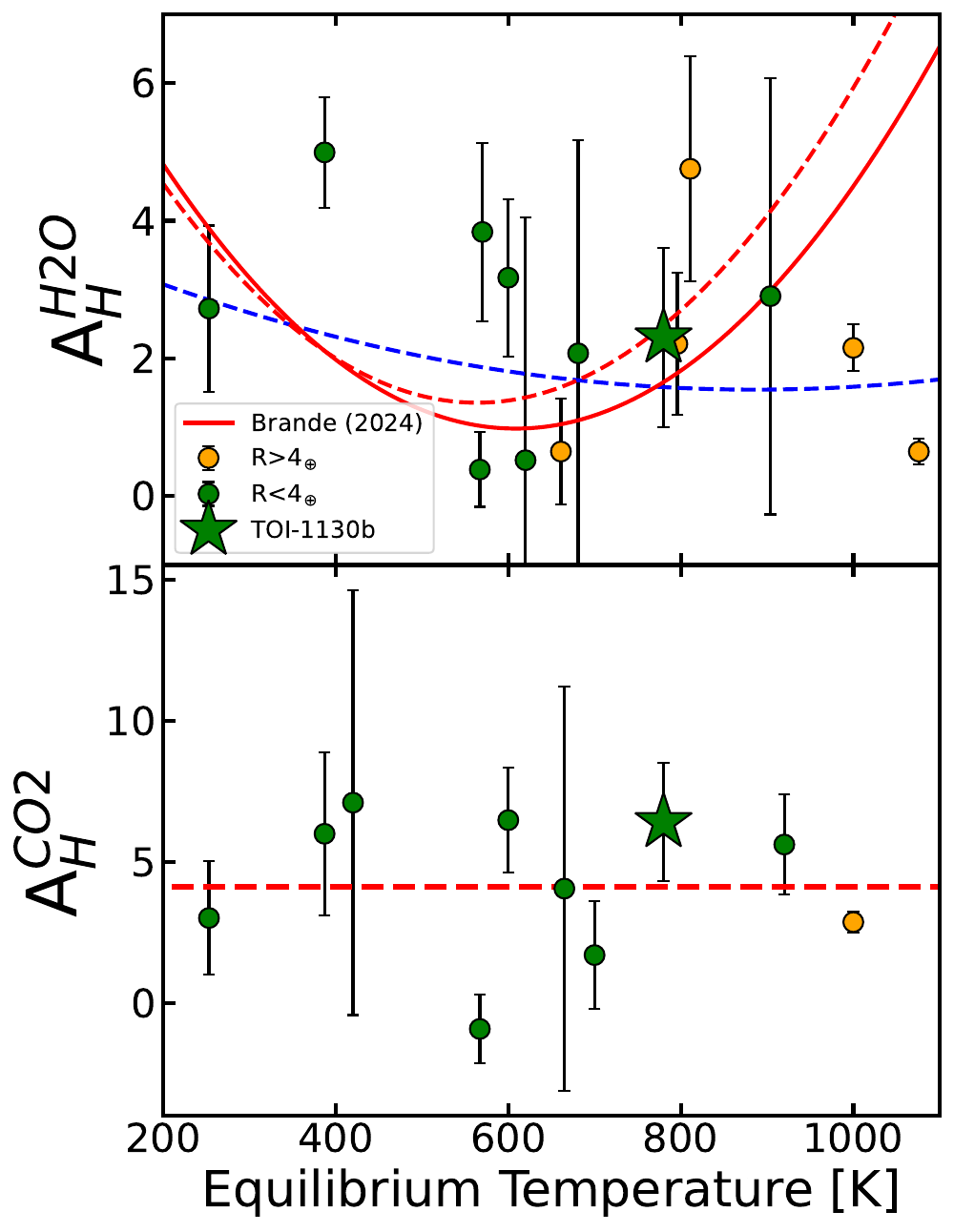}
    \caption{Scale height metric (Eqn\,\ref{eq:eqn1}) for mini/super-Neptunes as a function of temperature. Upper panel is water band $1.4\,\mu$m and lower panel CO$_2$ band $4.3\,\mu$m. We show TOI-1130b on this diagram with 
    $A_{H}^\mathrm{H_2O}=2.3\pm1.3$ and $A_{H}^\mathrm{CO_2}=6.4\pm2.1$ assuming a $\mu$ of 5.5. We overplot a tentative  quadratic trend reported in \citet{Brande2024} for $A_{H}^\mathrm{H_2O}$ using only HST data (red solid line). We refit a quadratic function with the JWST data points included and it gives a trend consistent with \citet{Brande2024} if we do not include data points beyond 1000~K (red solid line). Including the two points after 1000~K finds a weaker quadratic fit (blue dashed line). The flat dashed line in the lower panel shows the best-fit flat line to the sample. The temperatures used in this plot have been calculated assuming an albedo of 0.3 and full recirculation.
    The data used to generate this Figure is shown in Table\,\ref{tab:subneptune_final}.}
    \label{fig:AH_comparison}
\end{figure}

\section{Summary and Conclusions} \label{summary}

In this paper we present the transmission spectrum of TOI-1130b (3.66~R$_{\oplus}$, 19.8~M$_{\oplus}$, 825~K) observed using JWST NIRISS/SOSS and NIRSpec/G395H (one transit per instrument). We also combine the JWST observations with the TESS and CHEOPS transit depths (Section\,\ref{tess and cheops}) It is a part of a rare multi-transiting system, with an inner mini-Neptune (TOI-1130b) and an outer hot Jupiter (TOI-1130c). Such systems have a low occurrence rate of $\sim$7--8\% among the hot Jupiter population \citep{sha2026}, and cannot be formed through high-eccentricity migration \citep{dawson2018}. The two planets are in a 2:1 MMR \citep{borsato2024,korth2023}, which indicates formation beyond the ice line and migration driven by disk interactions \citep{goldreich1980}. Interestingly, we note that the inner mini-Neptune lies at the edge of the so called `radius cliff', which is a sparsely populated part of the exoplanet radius-period space \citep{Dattilo2023}. The unique architecture of this system constrains its formation history, and allows us to test whether formation location and atmospheric composition are correlated.

We detect H$_2$O (7.5$\sigma$), CO$_2$ (3.3$\sigma$) and SO$_2$ (3.6$\sigma$) and a tentative detection of CH$_4$ (2$\sigma$) from the combined transmission spectrum (Figure\,\ref{fig:full spectrum}). We used \texttt{PetitRadtrans} \citep{molliere_19} and \texttt{PLATON} \citep{zhang20} to run free chemistry retrievals (Section\,\ref{free retrieval}). We find that the molecular detections and detected abundances are consistent within 1$\sigma$ between the two independent retrieval frameworks (Table\,\ref{tab:table 1}). We ran equilibrium chemistry retrieval (Section\,\ref{equilibrium chemistry retrieval}) using \texttt{PetitRadtrans}. We constrained the atmospheric metallicity ($\log{Z/Z_{\odot}}=1.8_{-0.3}^{+0.4}$) and a 3$\sigma$ upper limit to the C/O ratio ($<$0.75). We post process our samples to calculate the atmospheric $\mu$ ($5.5^{+1.3}_{-0.8}$\,amu), confirming the volatile-rich nature of the atmosphere of this mini-Neptune. We run a grid of self-consistent forward models with \texttt{PICASO} \citep{2023Mukherjee_Picaso}, and compare with the observed transmission spectra of TOI-1130b. We find a best-fit metallicity of 100$\times$solar and a C/O ratio of 0.3, consistent with the results from the equilibrium chemistry retrieval. The detection of SO$_2$, which is a photochemical product \citep{tsai2023}, also confirms the high metallicity of the atmosphere of this planet. SO$_2$ has been shown to be strongly dependent on atmospheric metallicity \citep{crossfield2023}, and its detection rules out solar metallicity atmospheres \citep{crossfield2025,mukherjee2024}. We detect a slope in the optical part of the NIRISS/SOSS spectrum, which is consistent with the TESS and CHEOPS transit depths. The slope can be modeled using a Rayleigh/super-Rayleigh power-law model. However, haze models like Titan-like tholin, soot and phosphorus hazes can be ruled out for this atmosphere at $>$5$\sigma$ level of confidence (Figure\,\ref{haze models}). We did not detect a significant He absorption signal in NIRISS/SOSS transmission spectrum, and using \texttt{p-winds} models \citep{does_santos2022} we can put an upper limit on the atmospheric escape rate ($<$10$^{11}$g/s).

TOI-1130b joins two other mini-Neptunes, TOI-270d \citep{benneke2024} and GJ3470b \citep{beaty2024}, with volatile-rich high $\mu$ (5-6 amu) atmospheres. The $\mu$ for TOI-1130b falls in between a pure H/H$_e$ envelope and a pure steam envelope, and is consistent with an atmosphere where H/He and volatiles like H$_2$O are well mixed (miscible envelope mini-Neptune,  \citet{benneke2024}). If we assume that this planet was born in-situ, it would be difficult to explain its high volatile enrichment. At such close proximity to the star, the available solids are dominated by refractories. Although the gas phase can be enriched by migrating and evaporating pebbles, the presence of the hot Jupiter outside its orbit would cut-off this supply, the so-called `pebble-filtering' effect \citep{bitsch2021}. Alternatively, if the planet was born beyond the ice line, it would have access to volatile-rich solids. It has been hypothesized that mini-Neptunes could have originated beyond the water ice line and been born with volatile-rich atmospheres and interiors \citep{bitsch2019,aguichinne2021}. Therefore, the high $\mu$ we find for TOI-1130b is consistent with formation beyond the ice line and agrees with the inferred formation scenario from its orbital architecture. If we assume that TOI-1130b was born with a volatile-rich atmosphere, it is unlikely to have significant mass/radius evolution \citep{aguichine2025,rogers2025}. It has been shown that high $\mu$ atmospheres could have low atmospheric escape \citep{yoshida2025}, which is consistent with the non-detection of helium escape signal. Therefore, if we assume that this planet was born with a high $\mu$ atmosphere, it could have maintained its original position on the period-radius space at the end of its migration phase. This could potentially explain its location at the edge of the `radius cliff'. Furthermore, the detection of Carbon bearing molecules along with water is also consistent with a formation beyond the ice line, as in-situ formation models predict very low C/O ratios \citep{steinmeyer2026}. 

The discovery of H/He rich low mean molecular weight atmospheres of young sub-Neptunes \citep{barat2024b,thao2024,barat2025}, demographics of small planet population \citep{fulton2017} and young transiting planets \citep{vach2024}, evidence of atmospheric escape in mature mini-Neptunes \citep{orell2024} provide evidence for the `gas-dwarf' in-situ formation scenario for mini-Neptunes \citep{rogers2025}. However,  the volatile-rich atmosphere of TOI-1130b, along with its unique orbital architecture and position in the radius-period diagram provide evidence that the mini-Neptunes do not have a homogeneous formation history, and ex-situ volatile-rich formation mechanisms contribute to some fraction of their population \citep{burn2024,chakraborty2026}.

Alternatively, it has been suggested that the source of volatiles in the atmospheres of mini-Neptunes could come from interior-atmosphere interactions \citep{lichtenberg2025,dorn2021}. Although it has been shown to explain the volatile budget of low mass mini-Neptunes like TOI-270d \citep{nixon2025}, it is unclear whether such large scale magma oceans can exist for the more massive mini-Neptunes, like TOI-1130b \citep{breza2025,calder2025}. Further theoretical investigation of interior structure models of mini-Neptunes is necessary to understand the extent to which interior atmosphere interactions can alter primordial composition of high mass mini-Neptunes.

 These observations are associated with program GO 3385. SB acknowledges the support for this program provided by NASA through a grant from the Space Telescope Science Institute, which is operated by the Association of Universities for Research in Astronomy, Inc., under NASA contract NAS 5-03127. C.X.H acknowledges that her research is sponsored by the Australian Research Council Future Fellowship FT240100016. G.Z. acknowledges that his research is sponsored by the Australian Research Council.
 CHEOPS is an ESA mission in partnership with Switzerland with important contributions to the payload and the ground segment from
Austria, Belgium, France, Germany, Hungary, Italy, Portugal, Spain, Sweden,
and the United Kingdom. The CHEOPS Consortium would like to gratefully
acknowledge the support received by all the agencies, offices, universities, and
industries involved. Their flexibility and willingness to explore new approaches
were essential to the success of this mission. CHEOPS data analysed in this article will be made available in the CHEOPS mission archive. The authors acknowledge support from the Swiss NCCR PlanetS and the Swiss National Science Foundation. This work has been carried out within the framework of the NCCR PlanetS supported by the Swiss National Science Foundation under grants 51NF40182901 and 51NF40205606. J.K. acknowledges support from the Swedish Research Council (Project Grant 2017-04945 and 2022-04043) and of the Swiss National Science Foundation under grant number TMSGI2\_211697. H.P. acknowledges support by the Spanish Ministry of Science and Innovation with the Ram\'on y Cajal fellowship number RYC2021-031798-I. A.B. was supported by the Swedish National Space Agency. 
Funding from the University of La Laguna and the Spanish Ministry of Universities is acknowledged. ASTEP benefited from the support of the French and Italian polar agencies IPEV and PNRA in the framework of the Concordia station program.

%
%
\newpage
\bibliography{references.bib}{}
\bibliographystyle{aasjournal}
\appendix
\restartappendixnumbering

\section{JWST data reduction} \label{data reduction methods}

\subsection{NIRSpec data reduction} \label{NIRSpec_reduction}

We start our data analysis from the uncalibrated 4D images downloaded from MAST archive portal. We use the open source JWST data analysis pipeline \texttt{Eureka!}\ \citep{eureka} to analyze the images. In stage 1 standard data reduction steps such as superbias correction, dark current subtraction, jump detection during up-the-ramp sampling are applied using the standard \texttt{JWST} data reduction pipeline. We use a jump rejection threshold of 4$\sigma$. We test different values of jump rejection threshold (3-6$\sigma$) and we found that the ultimate white light curve showed the minimum Median Absolute Deviation (MAD) for 4$\sigma$. This step is effective in detecting cosmic rays. We also perform a crucial column-by-column median subtraction to account for the 1/f noise in the detector \citep{espinoza2023}. It has been noted by multiple JWST/NIRSpec G395H datasets where 1/f subtraction at the group level improves the final scatter in the white light curves by a factor of $\sim$2 \citep{alderson2023,May2023}. We mask the trace while estimating the background in this step.

We extract 1D stellar spectra from the Stage 2 outputs from \texttt{Eureka!}. We first use the Stage 3 extraction routine from \texttt{Eureka!}\ itself. It first locates the trace on each column and fits a Gaussian profile to it. It corrects for the curvature of the trace on the detector and aligns all the columns. Then it applies an optimal extraction \citep{horne1986}. A column-by-column median background subtraction is applied in this stage as well. To estimate the weights for the optimal extraction it constructs a median image which is smoothed spatially. Outlier rejection is applied both temporally and spatially to identify remnant cosmic rays, dead/hot pixels. The y and x position on the detector for both NRS1 and NRS2 is recorded during this stage. We do not find any signatures of mirror tilts/HGA moves during the visit from the y-position time series. All the integrations are centered on the same pixel with a MAD of 0.01 in both x and y direction. We use an aperture width of 8 pixels ensuring that the entire trace falls with the aperture window. Before the final light curve extraction we visually inspect the stacked stellar spectra to look for any additional `bad columns'. Such bad columns often appear as a vertical dark/bright line on the stacked stellar spectra. For the final light curve extraction these columns are masked.

\subsection{NIRISS SOSS data reduction}  \label{niriss_reduction}

The NIRISS SOSS data analysis was carried out from the uncalibrated files. We use the \texttt{exoTEDRF} pipeline \citep{radica2024} for the data reduction. We first proceed through the standard JWST data reduction steps such as superbias correction and reference pixel correction. We did not include the dark current subtraction step. The dark current is small and its reference file still contains residual 1/f noise \citep{radica2024} which is why it introduces more noise in the images. We tested this by running reductions with and without dark current subtraction step and found that including this step increases the scatter in the bluest part of the detector. For subsequent analysis we turn off the dark current subtraction step. In NIRISS/SOSS the background is wavelength dependent \citep{feinstein2023} and must be subtracted before the group level 1/f correction. The 1/f correction is applied using a column-by-column median subtraction and the subtracted background is re-added. Following this in Stage 1 ramp fitting is done. In this step abrupt jumps at the group level can be identified. 

In Stage 2, flat field subtraction is done, the background is removed and bad pixels are marked. We apply a PCA to the timeseries at this stage. We remove the first three components of the PCA to remove detector level effects. We find that removing more components did not decrease the MAD in the final white light curves. We note that a F227W image was not taken during these observations, so we skip the step where order 0 background contaminants are flagged. 

We use the Stage 2 outputs to extract the 1D stellar spectra. We use a box half-width of 17 pixels centered on the trace of order 1 and order 2 on the detector. The box width is chosen such that all of the trace falls within the aperture. We do not see any significant overlap between order 1 and 2 and therefore apply simple box extraction. Using the ATOCA algorithm \citep{radica2022} to extract the 1D spectra did not significantly affect the extracted spectra and scatter in the white light curves.


\section{Fitting JWST light curves} \label{subsec:lightcurve_analysis}

\subsection{NIRSpec light curve analysis}

 We model the transit light curves using a linear baseline and the transit is modeled using a \texttt{batman} model \citep{batman}. We fix the inclination, eccentricity and orbital period from \citet{borsato2024}. We fit for the mid transit time, semi-major axis, quadratic limb darkening coefficients, transit depth and a linear baseline. We used an MCMC sampler \texttt{emcee} \citep{emcee} to sample the parameter space with 50 walkers each taking 5000 steps. The first 1000 steps are discarded as burn-in. We find that the residuals in the white light curve shows some remaining red noise. 

The spectroscopic light curves are fit using two methods: same model as the white light curve, but we fix the mid transit time and semi-major axis from the respective white light curves. We also apply a common-mode (divide-white) method to detrend the spectroscopic light curves from wavelength independent systematics. The common-mode method is useful whenever systematics can be considered wavelength independent \citep{kreidberg14}. We let the linear slope free even for the common-mode corrected light curves. This is done because the NRS1 slope is known to be wavelength dependent \citep[e.g.\ see][]{sikora2024}. The quadratic limb darkening coefficients are fixed from \texttt{Exotic-LD} \citep{exotic_ld}. The stellar parameters are taken from \citet{korth2023}. The temperature of TOI-1130 (4350~K) is in the sweet spot such that fixing the limb darkening does not introduce significant bias in the retrieved transit depths \citep{espinoza2015}. We note that the residuals from the spectroscopic light curve fits do not show any signs of red noise and bin down as expected for white noise. We also note that the transmission spectra derived from both light curve fitting approaches yield consistent results (Figure\,\ref{NIRSpec_reduction}).


We also test the light curve fits by leaving mid transit time free. The spectrum is consistent with the fixed mid transit time case and the fitted mid transit times do not exhibit wavelength dependence. Recent observations of gas giants have shown that limb asymmetries can be quite common for these planets \citep{murphy2024,mukherjee2025}. One way to check for limb asymmetry is to fit the light curve using a symmetrical transit model like \texttt{batman}, but leave the mid transit time free. In case limb asymmetries are present they show wavelength dependent features in the mid transit time spectrum \citep{mukherjee2025}. However, given the SNR for a Neptune size planet is lower than a gas giant transit, we do not detect any limb asymmetry effect for this planet.

\begin{figure}
    \centering
    \includegraphics[width=1\linewidth]{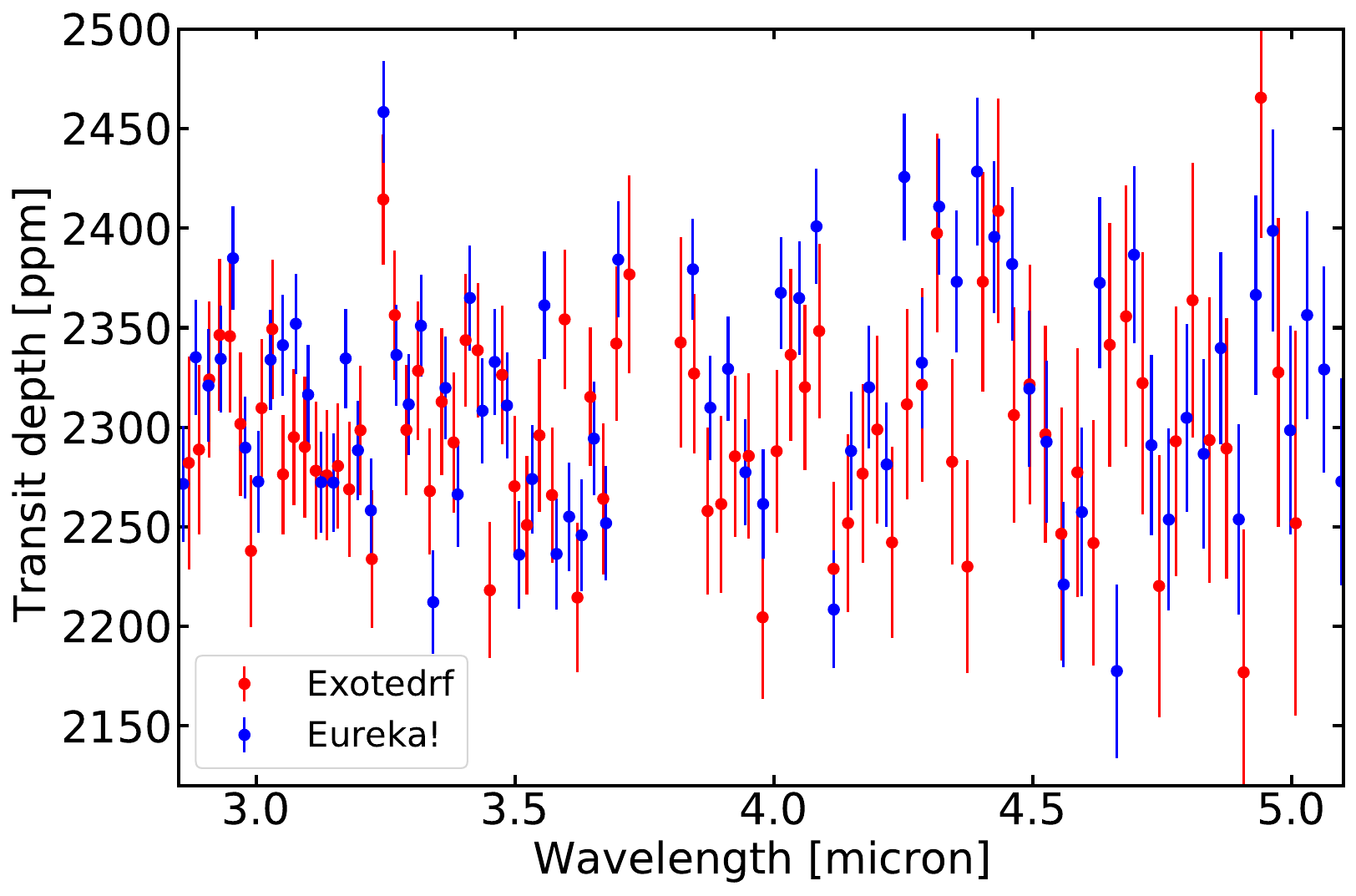}
    \caption{Comparison between NIRSpec G395H transmission spectra of TOI-1130b derived from independent data reductions from \texttt{Eureka!}\ and \texttt{exoTEDRF}. Light curve fits done using the common-mode method.}
    \label{fig:exoted_eureka}
\end{figure}

\begin{figure}
    \centering
    \includegraphics[width=1\linewidth]{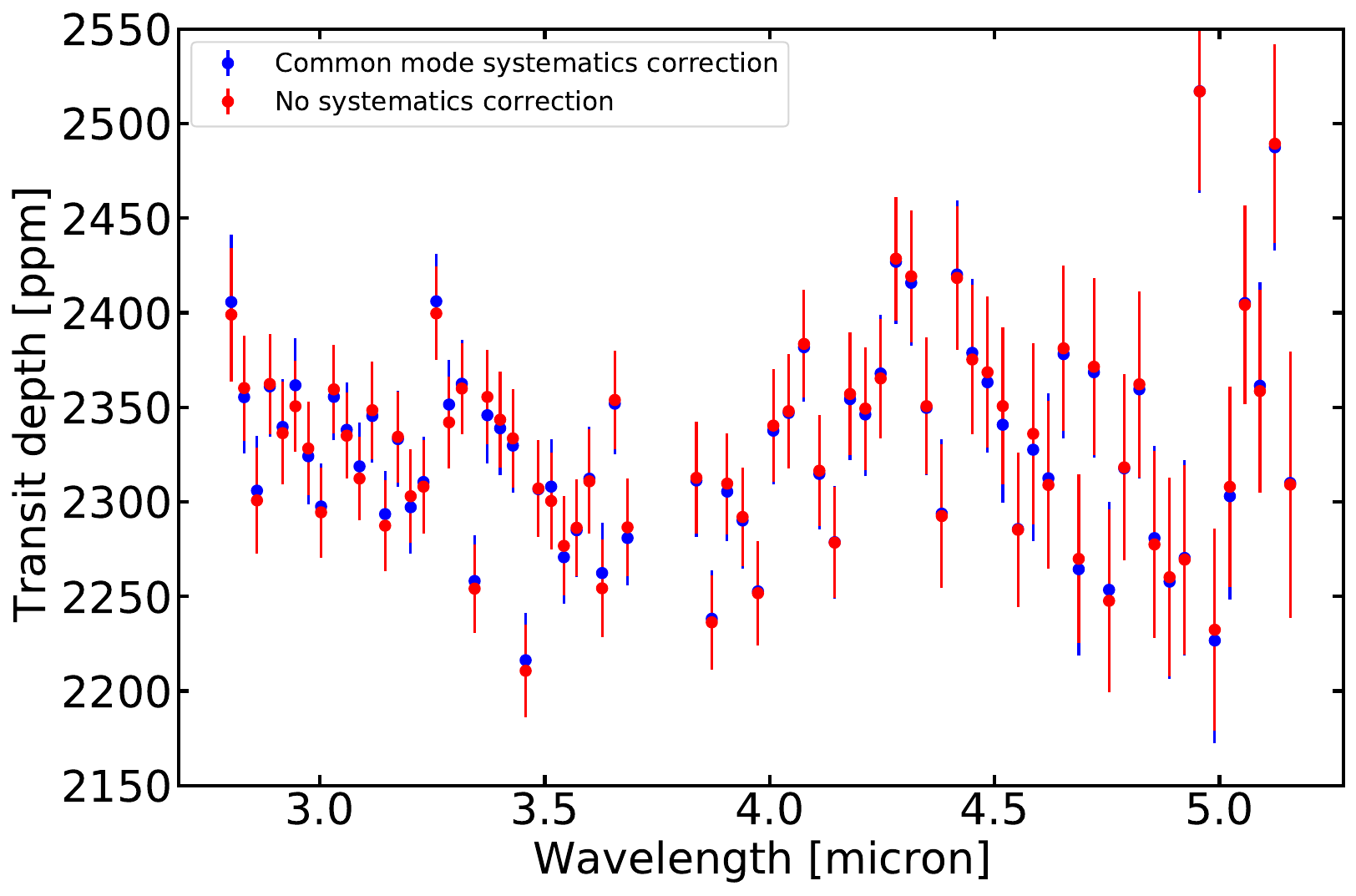}
    \caption{Comparison between NIRSpec G395H transmission spectra of TOI-1130b derived from independent light curve analysis approaches:  common-mode technique and no additional systematics model. Light analysis done on \texttt{Eureka!}\ reduction.}
    \label{fig:NIRSpec_common_mode_model_comparison}
\end{figure}

\subsection{NIRISS light curve analysis}

The NIRISS/SOSSS white light curves (Fig.\,\ref{fig:combined_white_lc}) were analyzed similarly compared to the NIRSpec G395H white light curves. The orbital parameters (inclination, eccentricity, orbital period) were fixed from literature values \citet{borsato2024}. The mid transit time, semi-major axis, quadratic limb darkening coefficients, transit depth and a linear baseline were free parameters in the white light curve fit. We used MCMC implemented through \texttt{emcee} \citep{emcee} to estimate the posterior distributions were these parameters. In the white light curve residuals we can see a `kink' around 60546.56 MJD, which is likely due to a mirror tilt/high gain antenna move. 

To account for this `kink' in the spectroscopic light curves, we take two approaches: first we remove all integrations before the `kink' to fit the spectroscopic light curves. Secondly, we apply a common-mode method where we derive a systematics model from the white light curve and by dividing the observed white light curve with the best-fit transit model. From the first fit we found that the linear slope of the baseline for both order 1 and order 2 do not show wavelength dependence. On visual inspection the detrended light curves from the common-mode correction do not show any residual features due to the `kink'. We also check that the spectroscopic light curves bin down consistently with white noise. Similar to the NIRSpec light curves, the quadratic limb darkening coefficients are fixed. The semi-major axis and mid transit times are fixed from the white light curve fits. We also run a fit where we let the mid transit time free for the spectroscopic light curves. We do not find significant wavelength dependence for the mid transit times, indicating that limb asymmetry effects are not significant in the NIRISS/SOSS bandpass.

\begin{figure}
    \centering
    \includegraphics[width=1\linewidth]{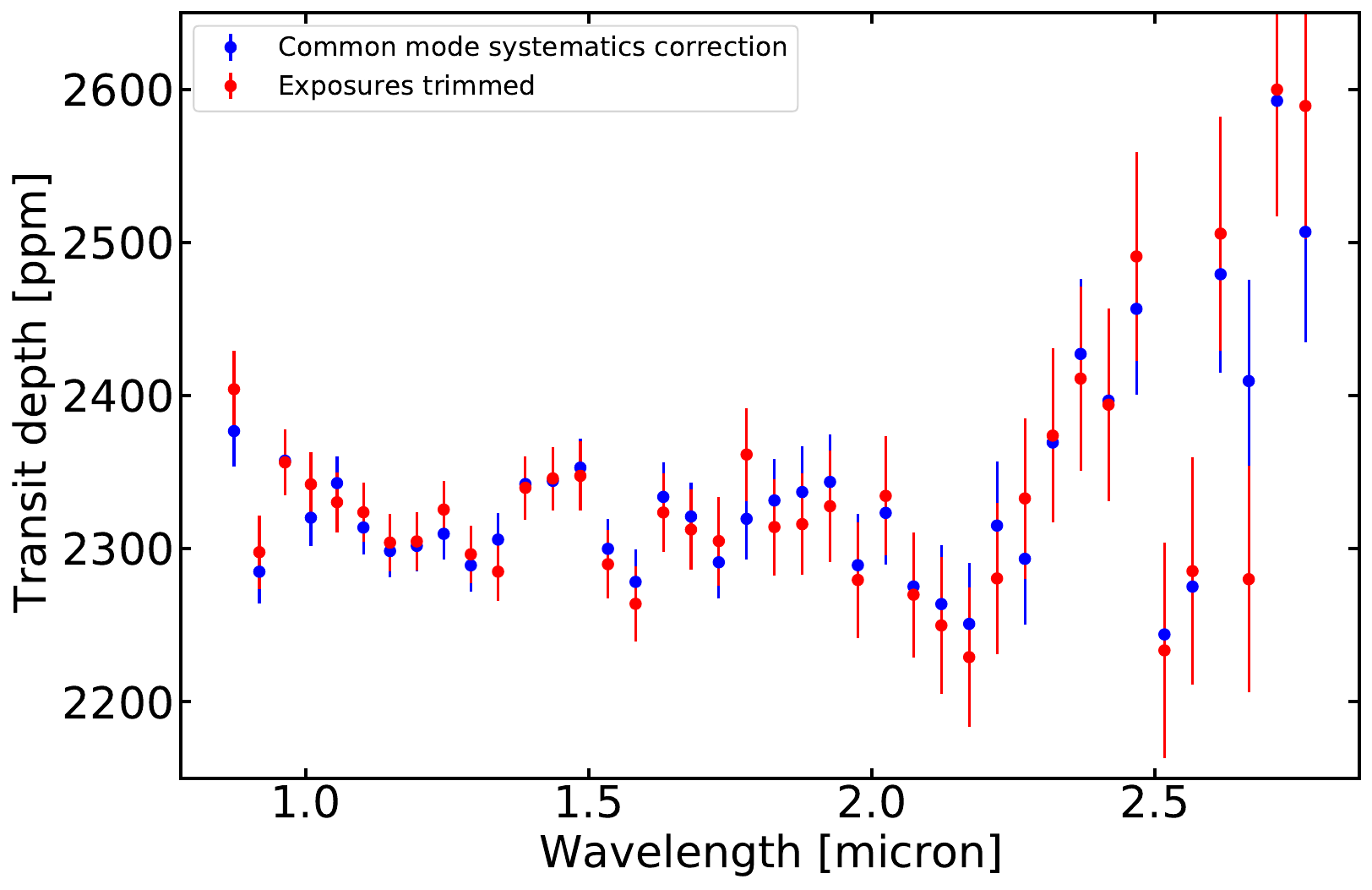}
    \caption{Comparison between NIRISS/SOSS transmission spectra of TOI-1130b derived from independent light curve analysis approaches:  common-mode technique and pre-kink exposures trimmed. Light analysis done on \texttt{exoTEDRF} reduction.}
    \label{fig:normal_common_mode_niriss_spectrum_comparison}
\end{figure}

A comparison between the spectrum derived from the brute force integration trimming approach and the common-mode correction method is shown in Figure\,\ref{fig:normal_common_mode_niriss_spectrum_comparison}. The extracted spectra and 1$\sigma$ uncertainties are consistent between the two light curve fitting approaches.

\subsection{Transmission spectrum binning}

The transmission spectra are derived by fitting the light curves at the native resolution. We derive the final transmission spectrum by binning the native resolution transmission spectra. To decide the binning, we calculate the autocorrelation function (ACF) on the native resolution transmission spectrum. For NIRISS order 1, order 2 and NRS2 the ACF is sharply peaked at 0, but for NRS1 the ACF has non 0 values till 0.015\,$\mu$m. We also calculate the Fourier transform of the native resolution transmission spectrum of NRS1 which shows a sharp peak corresponding to the 0.015\,$\mu$m. We choose a bin width of 0.03\,$\mu$m (50 pixels). This ensures that the Fourier transform of the bin has a 0 at the peak of native resolution spectrum Fourier transform. The other detectors do not show such features in the ACF or Fourier space. However, we still choose 50 pixel bins. Only for NIRISS SOSS order 2 we choose only 2 bins since the SNR is low for this order.

\begin{figure}
    \centering
    \includegraphics[width=1\linewidth]{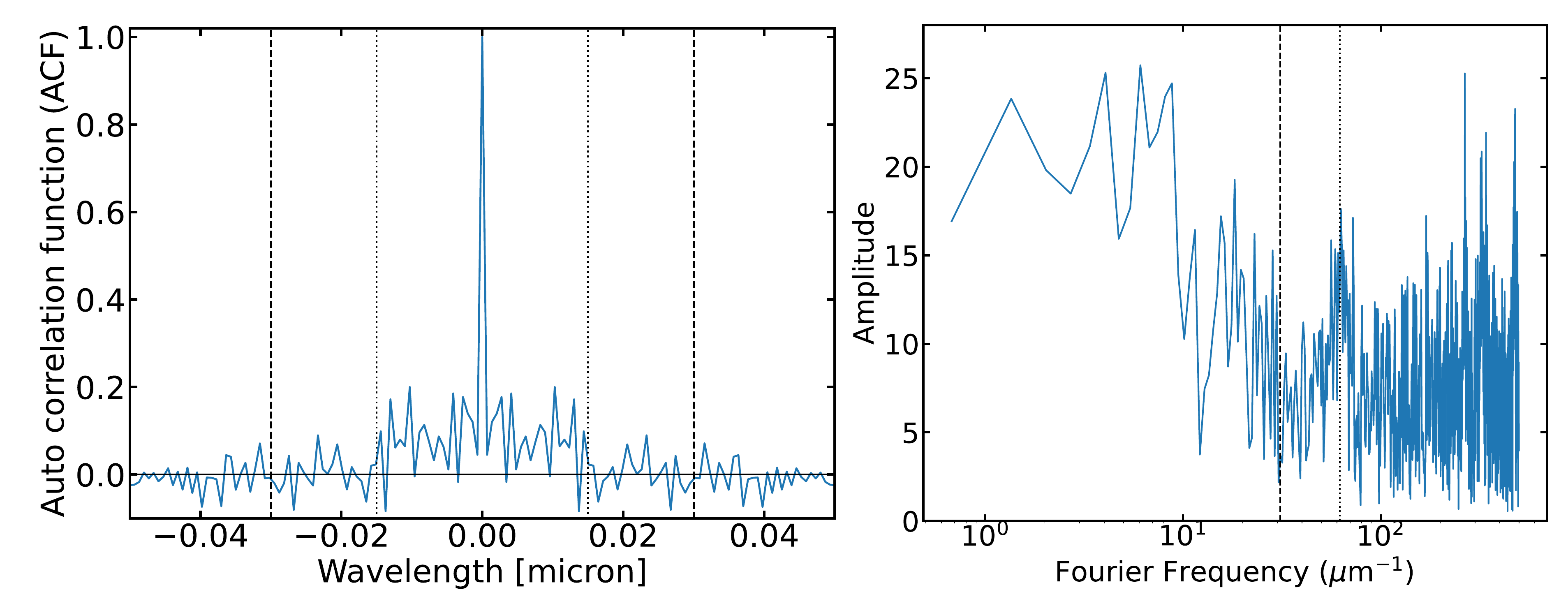}
    \caption{Autocorrelation function (left) and Fourier transform (right) of native resolution spectrum derived from common-mode correction of the NRS1 lightcurves. The ACF shows a non-zero values upto 0.015 microns (marked by dotted vertical line). The corresponding Fourier frequency also shows a peak (dotted line). We choose a bin width of 50 pixels which correspinds to twice the frequency of the peak. This ensures that Fourier transform of the binning function has a 0 at the peak.}
    \label{fig:nrs1_acf}
\end{figure}

\section{Atmospheric retrievals of TOI-1130b} \label{appendix:retrievals}

\subsection{PetitRadtrans free retrieval} \label{appendix:free retrieval}

\begin{figure}
    \centering
    \includegraphics[width=1\linewidth]{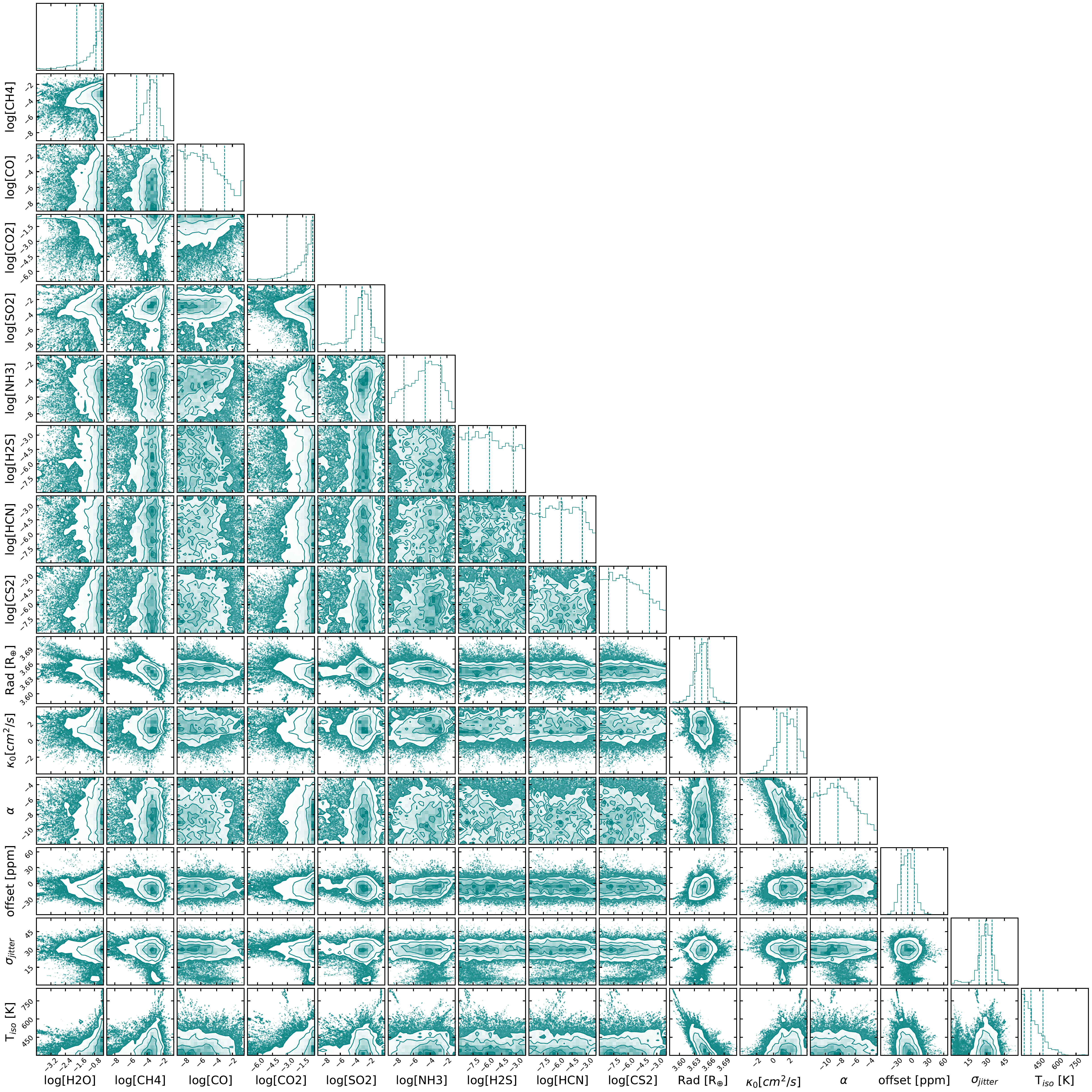}
    \caption{Posterior distribution from free chemistry retrieval (Section\,\ref{free retrieval}) on the combined TESS, CHEOPS and JWST transmission spectrum. The best-fit parameters are shown in Table\,\ref{tab:table 1}.}
    \label{fig:free posterior}
\end{figure}

We run a free chemistry retrieval on the transmission spectrum of TOI-1130b (Figure\,\ref{fig:full spectrum}). The retrieval is run on combined TESS/CHEOPS/NIRISS SOSS/NIRSpec G395H spectrum. The NIRISS and NIRSpec spectra are binned at the 50 pixel level. A plane parallel isothermal atmosphere is assumed. We choose a reference pressure of 0.01 bar. The radius at the reference pressure is a fitting parameter. The mass fraction of the included species is a free parameter. The mass of the planet is fixed to 19.8~M$_{\oplus}$ \citep{borsato2024}. A power-law model is assumed for the aerosol opacity. We put uniform priors on all the fitting parameters. For the molecule mass fractions we use bounds from -10 to -0.2 in log-space. We fit the transmission spectrum using an MCMC implemented through \texttt{emcee}. We use 50 walkers for 15000 steps, with the first 3000 treated as a burn-in. The retrieved posterior distribution from the free retrievals is shown in Figure\,\ref{fig:free posterior}.

We calculate the detection significances using a simple $\chi^{2}$ difference. The $\chi^{2}$ values are calculated for the best-fit model and a model where a particular molecule has been removed. Using $\chi^{2}$ tables we convert the difference in $\chi^{2}$ between the two models into a detection significance. In Table \ref{tab:detection significances} we show the $\chi^{2}$ values which were calculated for the different models.

We also calculate the Bayes factor to test the validity of estimated detection significances. Bayes factors have been considered a benchmark for detection significances recently in the community \citep{thorngren2026,constantinou2026,holmberg2023}. We approximate the Bayes factor from the posterior distribution samples using the Savage-Dickey approximation \citep{dickey1971}. This approximation applies only if the null model is nested within the full model and priors on nuisance parameters remains unchanged for the null model \citep{molotolov2016,mulder2020}. This approximation has been previously used for model comparisons in the context exoplanets and retrievals \citep{waldmann2015,kipping2024}.

\begin{figure}
    \centering
    \includegraphics[width=1\linewidth]{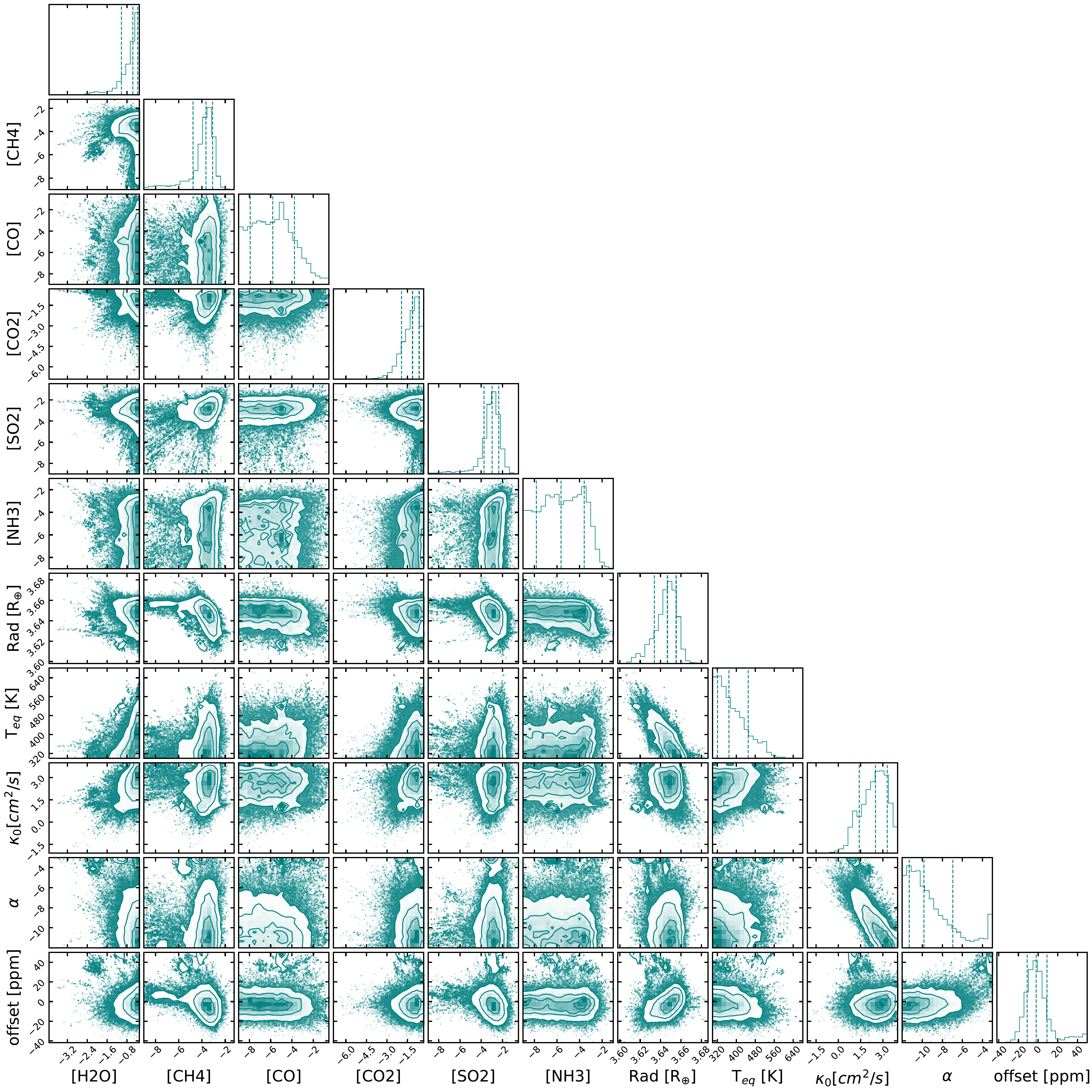}
    \caption{Posterior distribution from free chemistry retrieval (Section\,\ref{free retrieval}) on the combined TESS, CHEOPS and JWST transmission spectrum. We have removed the outliers between 2.3--2.7\,$\mu$m for this retrieval. The best-fit parameters are shown in Table\,\ref{tab:table 1}.}
    \label{fig:free_posterior_outlier_removed}
\end{figure}

\begin{figure}
    \centering
    \includegraphics[width=1\linewidth]{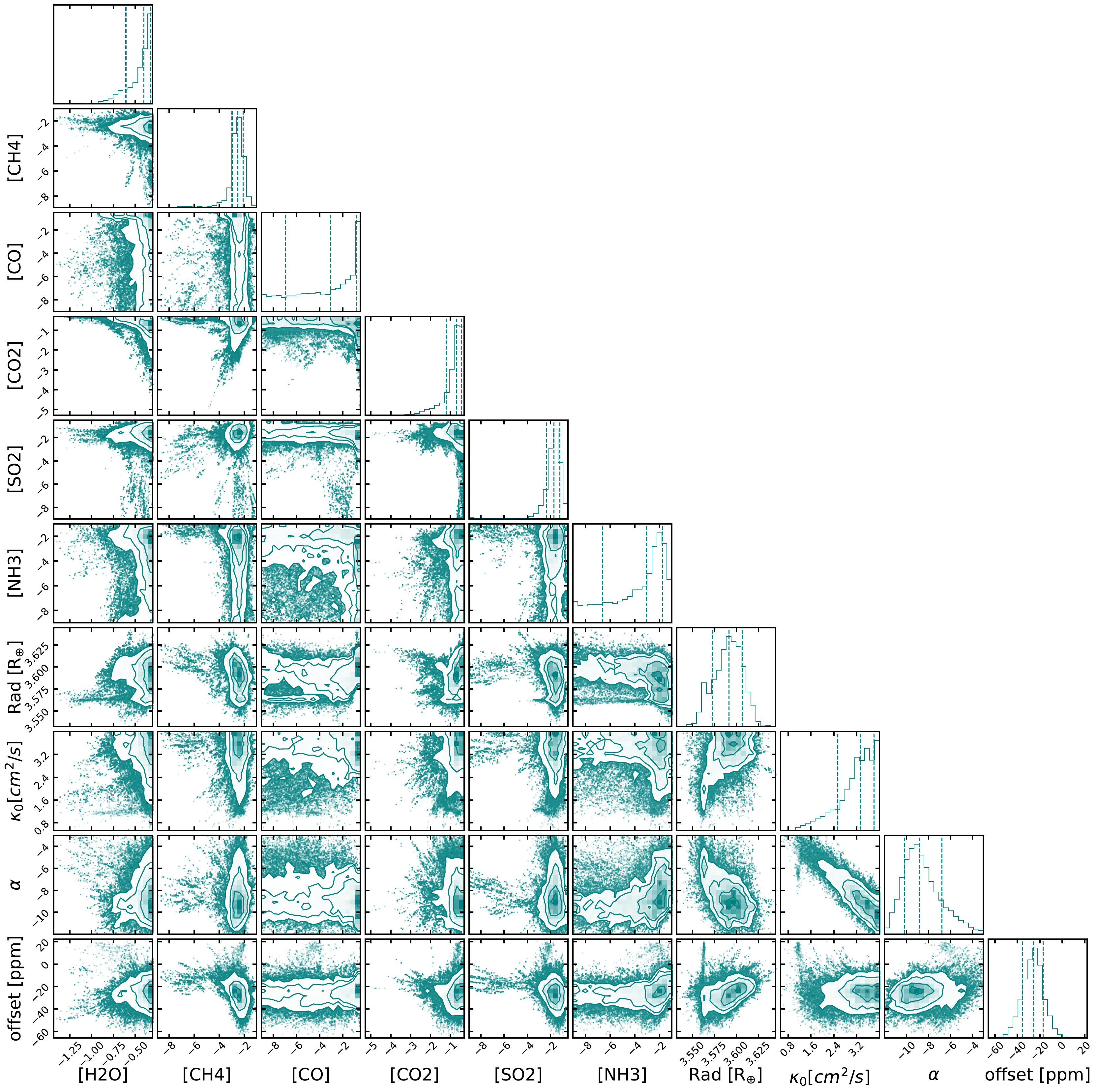}
    \caption{Posterior distribution from free chemistry retrieval (Section\,\ref{free retrieval}) on the combined TESS, CHEOPS and JWST transmission spectrum. We have fixed the isothermal temperature to 825~K \citep{borsato2024}. The best-fit parameters are shown in Table\,\ref{tab:table 1}.}
    \label{fig:free_ret_T_fixed}
\end{figure}

\subsection{PetitRadtrans equilibrium chemistry retrieval} \label{appendix:equilibrium retrieval}

We use the equilibrium chemistry module within \texttt{Petitradtrans} to calculate molecular abundances assuming equilibrium chemistry. It uses an equilibrium chemistry grid calculated using \texttt{EasyChem} \citep{molliere2017}. It takes atmospheric metallicity, C/O ratio, temperature and pressure as input parameters and computes the abundances over the entire pressure range. For this retrieval we have assumed isothermal atmosphere and included temperature as a free parameter. We also include metallicity and C/O ratio as free parameters. The results from the equilibrium chemistry retrievals are summarized in Table\,\ref{tab:table 2}.

\begin{figure}
    \centering
    \includegraphics[width=1\linewidth]{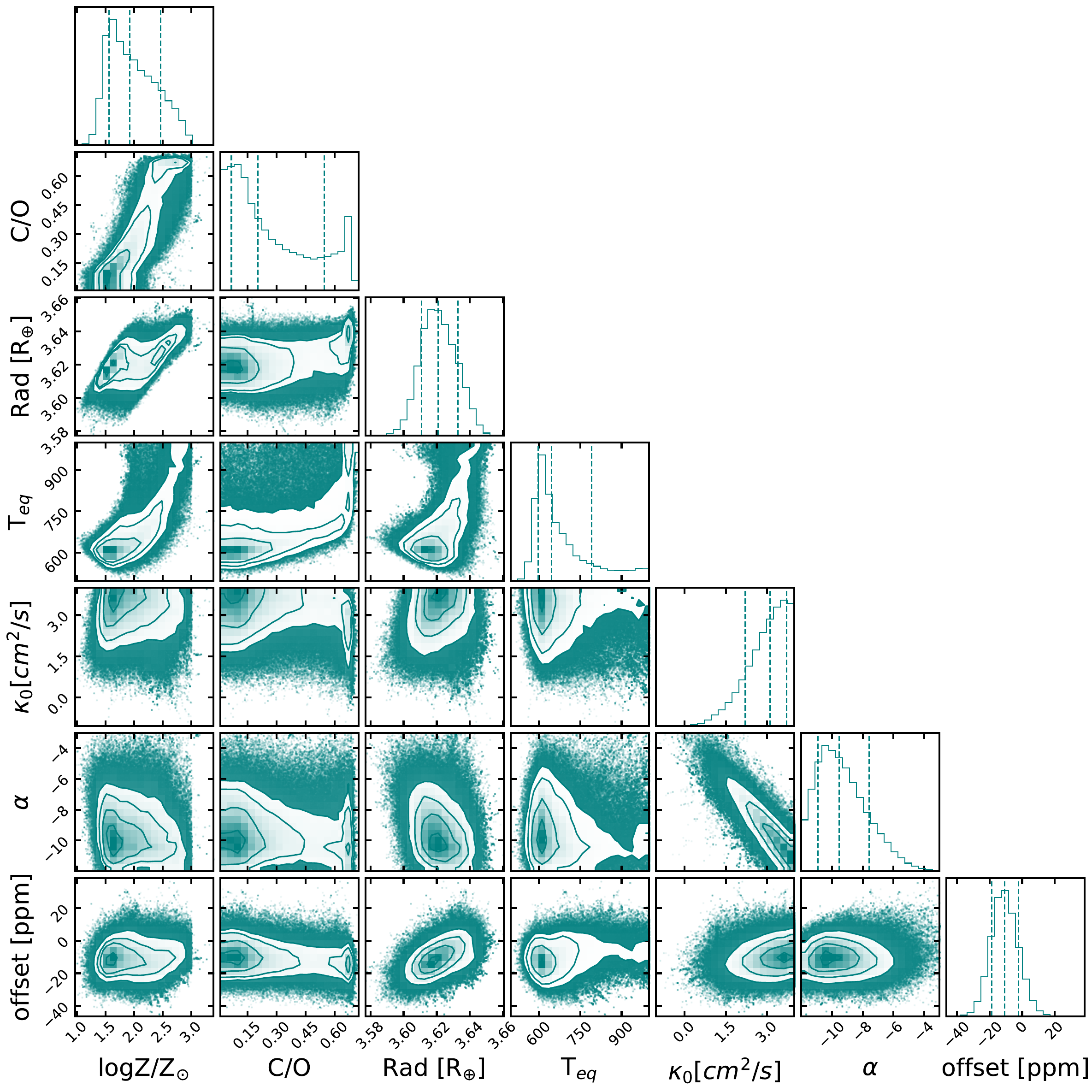}
    \caption{Posterior distribution from equilibrium chemistry retrieval (Section\,\ref{equilibrium chemistry retrieval}) on the combined TESS, CHEOPS and JWST transmission spectrum.}
    \label{fig:equilibrium posterior}
\end{figure}

\begin{figure}
    \centering
    \includegraphics[width=1\linewidth]{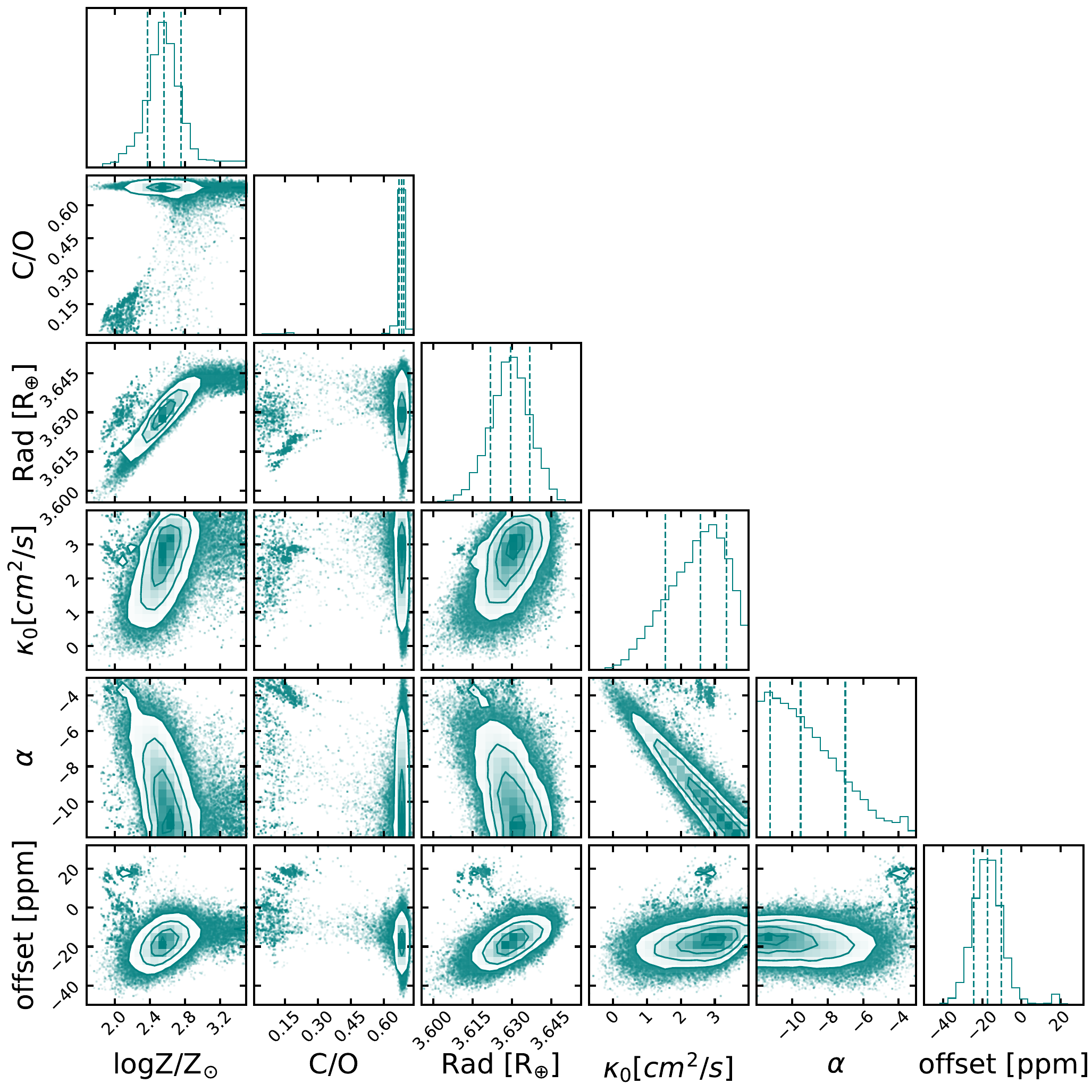}
    \caption{Posterior distribution from equilibrium chemistry retrieval (Section\,\ref{equilibrium chemistry retrieval}) on the combined TESS, CHEOPS and JWST transmission spectrum. Isothermal temperature fixed to 825~K.}
    \label{fig:equilibrium posterior_T800}
\end{figure}

\begin{figure}
    \centering
    \includegraphics[width=1\linewidth]{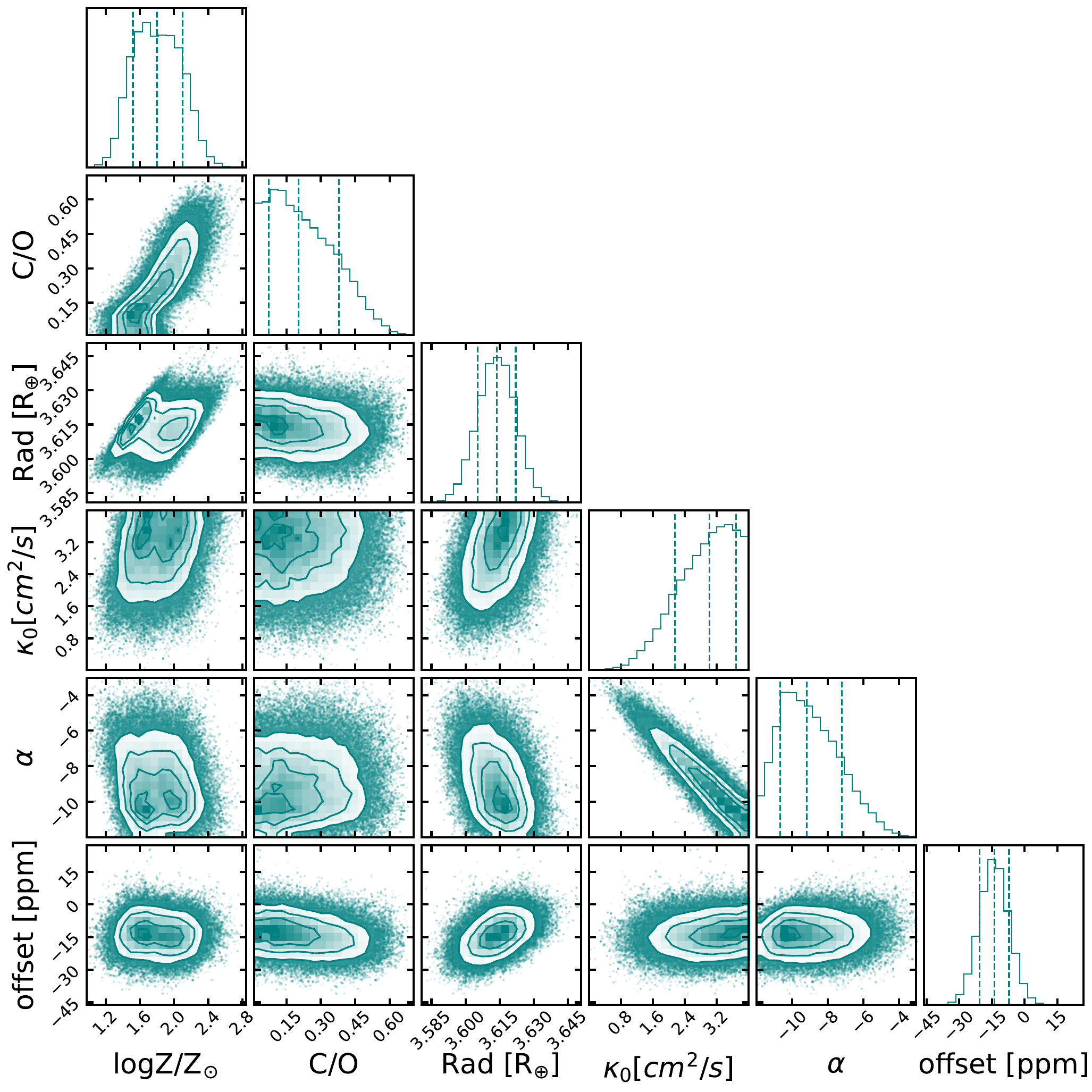}
    \caption{Posterior distribution from equilibrium chemistry retrieval (Section\,\ref{equilibrium chemistry retrieval}) on the combined TESS, CHEOPS and JWST transmission spectrum. Isothermal temperature fixed to 625~K.}
    \label{fig:equilibrium posterior_T600}
\end{figure}

\begin{figure}
    \centering
    \includegraphics[width=1\linewidth]{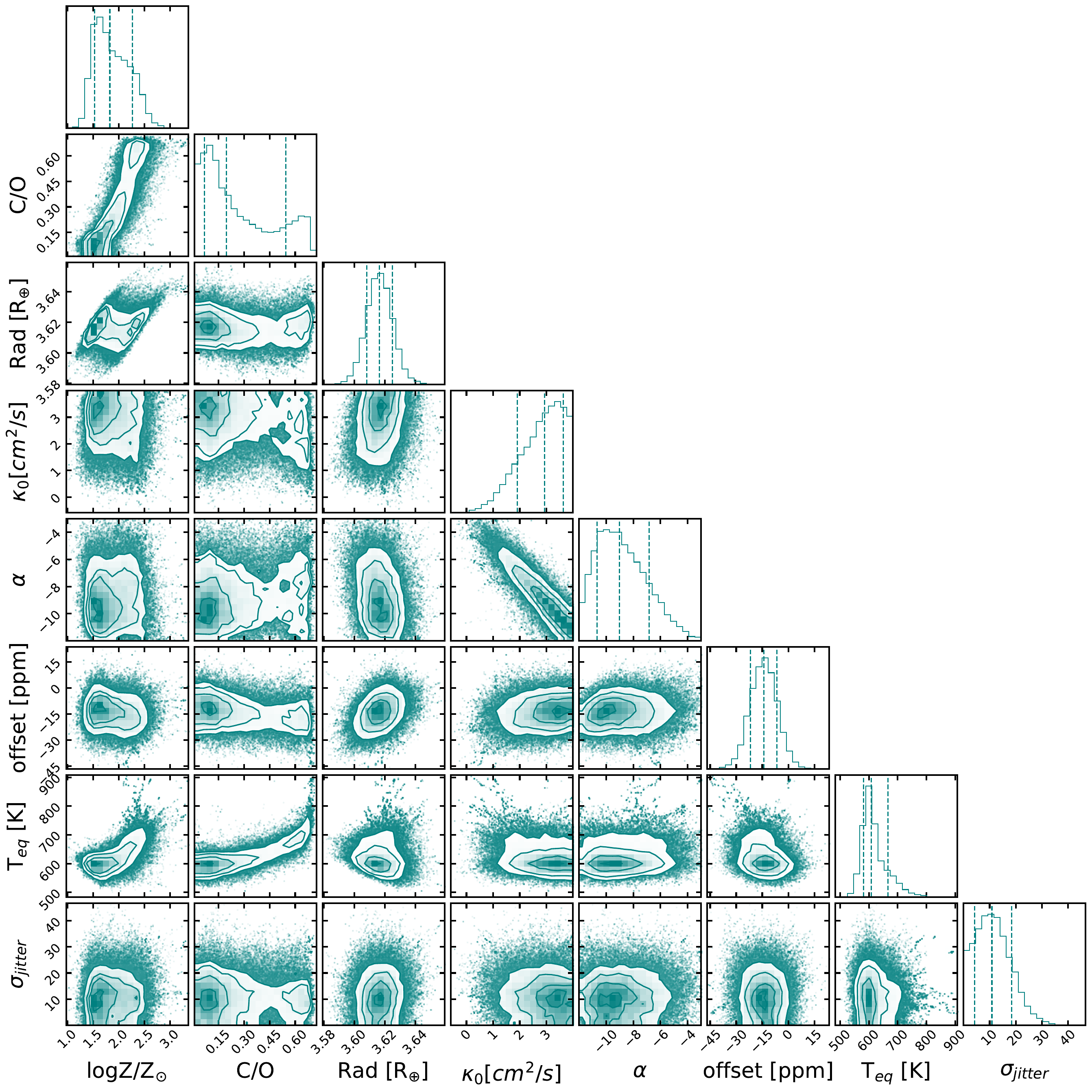}
    \caption{Posterior distribution from equilibrium chemistry retrieval (Section\,\ref{equilibrium chemistry retrieval}) on the combined TESS, CHEOPS and JWST transmission spectrum. We include a jitter term to account for residual red noise in the transmission spectrum.}
    \label{fig:equilibrium jitter}
\end{figure}

\section{Additional tables}

\begin{table*}
    \centering

    \begin{tabular}{c|c|c|c|c}
    \hline

       Planet  & Temperature [K] & Radius [R$_{\oplus}$] & $\mu$ [amu] & Reference\\
           \hline    \hline
       K2-18b  & 260 & 2.61 & 2.4$^{+1}_{-1}$ & \citet{madhusudhan2023,liu2025}\\
      TOI-270d   & 373 & 2.13 & 5.5$^{+1.2}_{-1.1}$ & \citep{benneke2024}\\
       TOI-776c  & 392 & 2.04 & $>$7 & \citep{teske2025}\\
        GJ1214b & 538 & 2.74 & $>$10 & \citep{kempton2023}\\
       GJ3470b  & 650 & 3.88 & 5.7$^{+2}_{-1.5}$ & \citep{beaty2024}\\
       GJ436b & 666 & 3.96 & $>$6 & \citep{knutson2014} \\
       TOI-836c & 626 & 2.58 & $>$6 & \citep{wallack2024} \\
       GJ3090 & 674 &  2.13 &$>$6 & \citep{ahrer2025} \\
       HAT-P-11b & 848 & 4.7 & 2.3$^{+0.3}_{-0.1}$ & \citep{chachan_2019} \\
       TOI-421b & 973 & 2.68 & 2.3$^{+1}_{-0.1}$ & \citep{davenport2025} \\
       HAT-P-26b & 970 & 6.3 & 2.5$\pm$0.3 & \citep{gressier2025} \\

    \hline
    \end{tabular}
    \caption{Table showing the radii, temperature and $\mu$ of the sample of planets used to generate Figure\,\ref{fig:mmw figure}. }
    \label{tab:MMW table}
\end{table*}

\begin{table*}[h]
\centering
\caption{Summary of data used to generate Figure\,\ref{fig:AH_comparison}. For planets where JWST observations are available we have used those. For others we have used HST observations. For planets where $\mu$ is not constrained we have assumed $\mu$=4 amu following \citet{Brande2024}.}
\resizebox{\textwidth}{!}{
\begin{tabular}{l|c|c|c|c|c|c|c|c|c|c|c|c}
\hline
Planet & $T_{\rm eq}$ [K] & $R_p$ [$R_\oplus$] & $R_\star$ [$R_\odot$] & $M_p$ [$M_\oplus$] & $\mu$ [amu] & $\Delta_{1.4}$ [ppm] & $\Delta_{1.2}$ [ppm] & err$_{1.4}$ [ppm] & $\Delta_{4.3}$ [ppm] & $\Delta_{4.1}$ [ppm] & err$_{4.1}$ [ppm] & Reference \\
\hline
K2-18b         & 260  & 2.61 & 0.44 & 8.80  & 2.4   & 2945 & 2878 & 21 & 2960 & 2886 & 35 & \citet{madhusudhan2023} \\
TOI-270d       & 373  & 2.13 & 0.38 & 4.78  & 5.5   & 2935 & 2825 & 13 & 3002 & 2870 & 45 & \citet{benneke2024} \\
TOI-776c       & 392  & 2.05 & 0.54 & 7.00  & 7.0   & - & - & - & 1180 & 1140 & 30 & \citet{teske2025} \\
GJ~1214b       & 538  & 2.74 & 0.21 & 8.40  & 10.0  & 13522 & 13495 & 27 & 135056 & 135121 & 60 & \citet{kempton2023} \\
HD~3167c       & 546  & 3.01 & 0.86 & 9.80  & 4.0   & 955 & 909 & 11 & - & - & - & \citet{mikal-evans2021} \\
GJ~3470b       & 650 & 3.88 & 0.48 & 11.2 & 5.7 & 6122 & 5953 & 43 & 6109 & 5764 & 70 & \citet{beaty2024} \\
GJ~436b        & 666 & 3.96 & 0.46 & 22.0 & 6.0 & 7036 & 7020 & 77 & 0 & 0 & 0 & \citet{knutson2014} \\
TOI-674b       & 676  & 5.25 & 0.42 & 23.00 & 4.0   & 13195 & 13110 & 72 & - & - & - & \citet{brande2022} \\
TOI-836c       & 626  & 2.58 & 0.67 & 9.60  & 6.0   & - & - & - & 1220 & 1180 & 50 & \citet{wallack2024} \\
HD~97658b      & 729  & 2.12 & 0.73 & 8.30  & 4.0   & 923 & 906 & 18 & - & - & - & \citet{knutson2014b} \\
GJ~3090b       & 674  & 2.13 & 0.51 & 3.34  & 5.9   & - & - & - & 1540 & 1490 & 40 & \citet{ahrer2025} \\
HAT-P-11b      & 848  & 4.70 & 0.75 & 25.00 & 2.4   & 3498 & 3377 & 40 & - & - & - & \citet{chachan_2019} \\
HD~106315c     & 1129  & 4.35 & 1.30 & 15.00 & 4.0   & 1099 & 1029 & 17 & - & - & - & \citet{kreidberg2022} \\
HIP~41378b     & 985  & 2.90 & 1.40 & 8.30  & 4.0   & 395 & 373 & 17 & - & - & - & \citet{edwards2022} \\
TOI-421b       & 973  & 2.68 & 0.87 & 7.00  & 2.33  & - & - & - & 820 & 640 & 40 & \citet{davenport2025} \\
HAT-P-26b      & 970 & 6.30 & 0.79 & 19.00 & 2.5   & 5318 & 4912 & 45 & 5790 & 5250 & 50 & \citet{gressier2025} \\
HD~219666b     & 1048 & 4.89 & 0.40 & 17.60 & 4.0   & 1833 & 1672 & 33 & - & - & - & \citet{murphy2025} \\
\hline

\end{tabular}
}
\label{tab:subneptune_final}
\end{table*}

\section{Log of additional observations of TOI-1130b}
\label{appendix:observations}

We listed all TOI-1130b observations used in addition to \citet{borsato2024} and the JWST observations in the Table below. 

\begin{table}
    \centering

    \begin{tabular}{c|c|c|c|c|c|c}
        \hline
        \hline
        $\#$ & & DATA ID & Start date & Duration & Frames & exp. time   \\
        & & & (UTC) & (h) & & (s)  \\
        \hline
        1 & CHEOPS &  CH\_PR149003\_TG000401\_V0300 &  2024/08/24 & 10.86 & 421  &  60 \\
       2  & TESS & tess2025180145000-s0094 & 2025/06/29 & 605.76 & 109036 &  20   \\
       3  & MINERVA & TOI-1130~b  & 2023/05/12  & 4.98 & 265 & 66\\
       5  & LCOGT &  TIC254113311-02\_20230517\_LCO-SSO-1m0\_ip  & 2023/05/17  & 6.9 & 543 & 15\\
     6    & LCOGT &  TIC254113311-02\_20230606\_LCO-SAAO-1m0\_ip & 2023/06/06  & 6.65 & 529 & 15\\
   7    & LCOGT &  TIC254113311-02\_20230610\_LCO-SAAO-1m0\_ip & 2023/06/10  & 5.58 & 448 & 45\\
       \hline

        \hline
    \end{tabular}
    \caption{Additional TOI-1130b transits used to constraint the transit timing model of TOI-1130 system. The CHEOPS and TESS data are also included in the global fit to provide the optical transit depth for the TOI-1130b spectra.}
    \label{tab:transit_observations}
\end{table}

\end{document}